\documentclass[aps,prl,twocolumn,superscriptaddress,longbibliography,10pt]{revtex4-1}
\usepackage[bookmarks=true,colorlinks,citecolor=blue,urlcolor=blue]{hyperref}
\usepackage{ucs}
\usepackage[latin1]{inputenc}
\usepackage[english]{babel}
\usepackage[usenames, dvipsnames]{xcolor}
\usepackage[export]{adjustbox}
\usepackage{amsmath, 
			amsfonts,
			booktabs,
            microtype,
            mathtools, 
            graphicx,
            color,
            soul,
            nicefrac,
            float,
            url, 
            placeins,	
			braket,
			appendix} 	
\usepackage{pdfpages}

\makeatletter
\AtBeginDocument{\let\LS@rot\@undefined}
\makeatother

\DeclareRobustCommand{\mb}[1]{%
  \ifmmode\text{\hl{$#1$}}\else\hl{#1}\fi
}

\begin{document}
\title{Unitary and non-unitary quantum cellular automata with Rydberg arrays}

\author{T. M. Wintermantel}\affiliation{Physikalisches Institut, Universit\"at Heidelberg, Im Neuenheimer Feld 226, 69120 Heidelberg, Germany}\affiliation{ISIS (UMR 7006) and IPCMS (UMR 7504), University of Strasbourg and CNRS, 67000 Strasbourg, France}
\author{Y. Wang}\affiliation{ISIS (UMR 7006) and IPCMS (UMR 7504), University of Strasbourg and CNRS, 67000 Strasbourg, France}
\author{G. Lochead}\affiliation{ISIS (UMR 7006) and IPCMS (UMR 7504), University of Strasbourg and CNRS, 67000 Strasbourg, France}
\author{S. Shevate}\affiliation{ISIS (UMR 7006) and IPCMS (UMR 7504), University of Strasbourg and CNRS, 67000 Strasbourg, France}
\author{G. K. Brennen}\affiliation{Center for Engineered Quantum Systems, Dept. of Physics \& Astronomy, Macquarie University, 2109 NSW, Australia}
\author{S. Whitlock}\email{whitlock@unistra.fr}\affiliation{ISIS (UMR 7006) and IPCMS (UMR 7504), University of Strasbourg and CNRS, 67000 Strasbourg, France}
\date{\today}

\begin{abstract}
We propose a physical realization of quantum cellular automata (QCA) using arrays of ultracold atoms excited to Rydberg states. The key ingredient is the use of programmable multifrequency couplings which  generalize the Rydberg blockade and facilitation effects to a broader set of non-additive, unitary and non-unitary (dissipative) conditional interactions. Focusing on a 1D array we define a set of elementary QCA rules that generate complex and varied quantum dynamical behavior. Finally we demonstrate theoretically that Rydberg QCA is ideally suited for variational quantum optimization protocols and quantum state engineering by finding parameters that generate highly entangled states as the steady state of the quantum dynamics.
\end{abstract}

\maketitle
Today there exists a wide variety of viable physical platforms for quantum information processing (QIP), including ultracold atoms, ions, impurities, photons and superconducting circuits. Each platform has its own unique advantages (and challenges) concerning important qualities such as isolation from the environment, qubit coherence time, gate speeds, scalability, addressability and interaction control. Therefore, to bring important and classically intractable problems within reach, protocols for quantum information processing must be robust and highly optimized to exploit the particular advantages of continually improving quantum hardware~\cite{Preskill2018}.

\begin{figure}[bt]
	\includegraphics[width=1.\columnwidth]{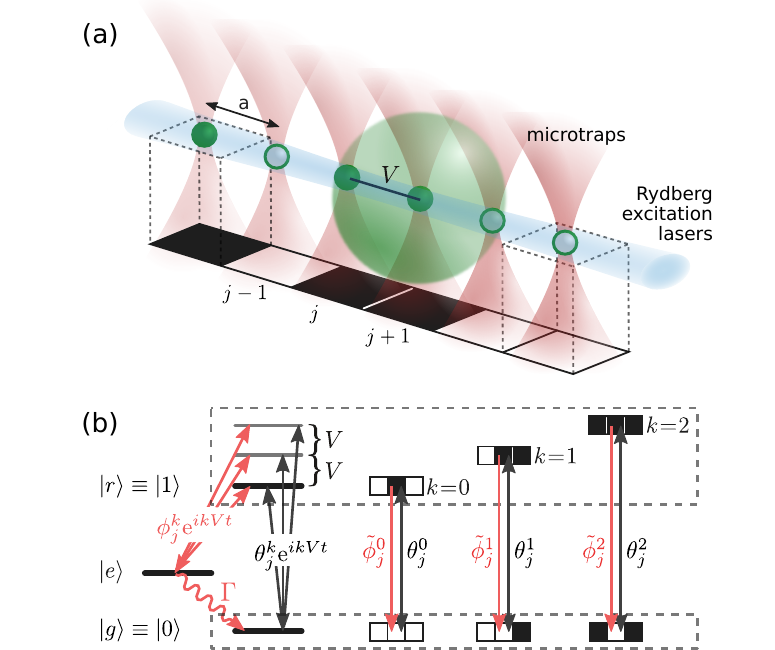}
	\caption{Physical platform for quantum cellular automata based on arrays of Rydberg atoms. (a) Proposed setup showing a 1D array of atoms held in optical microtraps with period $a$ and nearest-neighbor Rydberg-Rydberg interaction strength $V$. (b) Each atom can be described as a three-state system: $|g\rangle$ (open symbols), $|r\rangle$ (solid symbols) and an additional short lived state $|e\rangle$. The $|g\rangle \leftrightarrow |r\rangle$ and $|r\rangle \leftrightarrow |e\rangle$ transitions on site $j$ are coupled by multifrequency fields with detunings $kV$ and coupling strengths $\theta^k_j$ and $\phi^k_j$ respectively. This system can be reduced to an effective two-state system with programmable $K$-body interactions (shown on the right for $K=3$, see text for details), where the couplings $\theta^k_j$ and $\tilde{\phi}^k_j$ realize unitary (reversible) and non-unitary (dissipative) conditional interactions dependent on the number of excited neighbors $k$.}
	\label{fig:QCA_setup}
\end{figure}

One promising platform for QIP is based on trapped ultracold Rydberg atoms~\cite{Lukin2001,saffman2010,weimer2010,saffman2016}. Their distinguishing features include: (i) the availability of fast and switchable multiqubit interactions~\cite{Lukin2001,Unanyan2002,weimer2010,Isenhower2011,Ebert2015,Zeiher2015,gambetta2019} and (ii) the possibility for non-trivial dissipative interactions, which rather than destroying entanglement can actually enhance and protect it~\cite{verstraete2009,weimer2010,rao2013,Carr2013,rao2014,Shao2014,Roghani2018,Ding2019}. 

In this letter, we propose a physical implementation of the quantum cellular automata (QCA) paradigm~\cite{Wiesner2009,farrelly2019,arrighi2019} based on Rydberg atoms. This opens up an approach to QIP which is inherently parallelizable, does not require individual addressing of each qubit~\cite{lloyd1993,watrous1995one,raussendorf2005,Benjamin2000,Benjamin2001,twamley2003,Brennen2003, vollbrecht2006reversible} and takes full advantage of both unitary and non-unitary multiqubit interactions, potentially providing a viable and computationally universal alternative to gate-based~\cite{saffman2010, weimer2010,Petrosyan_2016,ostmann2017} and quantum adiabatic protocols~\cite{Keating2013,Lechner2015,glaetzle2017}. The key idea is to use programmable multifrequency excitation and depumping of Rydberg states that implements a set of conditional interactions in analogy with classical cellular automata. We show that this leads to a rich diversity of controllable quantum dynamics in both discrete and continuous time evolution. Finally, we numerically demonstrate a powerful approach for generating highly entangled quantum states by embedding Rydberg QCA within a variational quantum optimization loop~\cite{Moll2018,kokail2019}. 

\noindent\textit{Physical system:-} As a physical platform we consider an array of three-level systems consisting of a ground state $|g\rangle\equiv\ket{0}$, a strongly-interacting (Rydberg) state $|r\rangle\equiv\ket{1}$ and a short-lived intermediate state $|e\rangle$ used to mediate non-unitary interactions shown in Fig.~\ref{fig:QCA_setup}. This could be realized for example using single atoms~\cite{Schauss2012,labuhn2016,bernien2017,Sanchez2018,Lienhard2018}, trapped ions~\cite{Muller_2008,gambetta2019,zhang2019} or Rydberg blockaded atomic ensembles~\cite{Ebert2015,Whitlock_2017,Letscher_2017}. For simplicity we consider an equidistant 1D chain of trapped atoms restricted to nearest neighbor interactions $V$.  Two fields consisting of several discrete frequency components couple the $|g\rangle\leftrightarrow |r\rangle$ transition and the $|r\rangle\leftrightarrow |e\rangle$ transition, depicted in Fig.~\ref{fig:QCA_setup}(b). Within the rotating wave approximation the system is described by a time-dependent quantum master equation in Lindblad form ($\hbar=1$): $ \partial_t \rho = \mathcal L[\rho] = -i[\hat H,\rho] + \mathcal{D}[\rho]$, where
\begin{align}\label{eq:fullhamiltonian}
\hat{H}\!&=\!\sum_{j,k} \!\Bigl( \frac{\theta_j^k}{2}e^{ikVt}  \hat \sigma_j^{gr}\!+\! \frac{\phi^k_j}{2}e^{ikVt}\hat \sigma_j^{er}\! + h.c.\Bigr )\!+\!V \hat \sigma_j^{rr} \hat \sigma_{j+1}^{rr}, 
\end{align}
defining $\hat \sigma^{ab}_j=|a\rangle\langle b|$ acting on site $j$ and the nearest neighbor interaction strength $V$. The time dependent phase factors describe discrete components of the multifrequency fields with detunings $kV$ ($k=\{0,1,2\}$) and coupling strengths $\theta^k_j,\phi^k_j$. In the following we allow these couplings to be site dependent (e.g. applied independently to even and odd sites), but they can also be uniform for the whole system. Dissipation is included via the term $\mathcal{D}[\rho]=\sum_j \hat L_j \rho \hat L_j^\dagger -(\hat L_j^\dagger \hat L_j\rho+\rho \hat L_j^\dagger \hat L_j)/2$ where we define the jump operators $\hat L_j = \sqrt{\Gamma}\hat\sigma^{ge}_j$ describing spontaneous decay out of the $|e\rangle$ state. Rydberg state decay $\sqrt{\gamma}\hat \sigma^{gr}_j$ is assumed to be much slower than the rest of the dynamics and will be neglected for the moment. 

In the limit $V\gg \Gamma > \theta^k_j,\phi^k_j$ one can reduce the full quantum master equation to an effective two-level system (i.e. $|0\rangle$,$|1\rangle$) with time-independent $3-$body conditional interactions (see Supplemental Material for the full derivation). Briefly, we transform the Hamiltonian \eqref{eq:fullhamiltonian} to an interaction picture with respect to the nearest neighbor Rydberg-Rydberg interactions~\cite{Lesanovsky2011} and then adiabatically eliminate the time-dependent phase factors using a large frequency expansion~\cite{Sanz2016, james2007effective}. In a second approximation we adiabatically eliminate the rapidly decaying $|e\rangle$ states using the effective operator formalism~\cite{Reiter2012}, yielding an effective time-independent master equation defined by
\begin{align}\label{eq:PXP1}
\hat H^\mathrm{eff} &= \frac{1}{2}\sum_{j} \sum_{\alpha,\beta}\theta^k_{j} \mathrm{P}_{j-1}^\alpha \hat X_{j} \mathrm{P}_{j+1}^\beta,\\
\hat L^\mathrm{eff} &= \frac{1}{2}\sum_{j} \sum_{\alpha,\beta}\sqrt{\tilde \phi^k_j} \mathrm{P}_{j-1}^\alpha (\hat X_{j}-i \hat Y_{j}) \mathrm{P}_{j+1}^\beta,\label{eq:PXP2}
\end{align}
where $\sqrt{\tilde \phi^k_j} \approx \phi^k_j/\sqrt{\Gamma}$ and assuming $\theta^k_j \in \mathbb{R}$. The double sum over $\alpha,\beta$ goes from $0$ to $1$ with $k=\alpha+\beta$, $\mathrm{P}^\alpha=|\alpha\rangle\langle \alpha|$ and $\hat X_{j},\hat Y_{j}$ (and $\hat Z_j$) are Pauli matrices. 
Higher order corrections to this model enter as effective level shifts and couplings $\propto |\theta^k_j|^2/V,|\phi^k_j|^2/V$ which can be mostly neglected for experimentally relevant parameters. See Supplemental Material for benchmarking of the effective model Eqs.~\eqref{eq:PXP1} and \eqref{eq:PXP2} against the full time-dependent three-level model Eq.~\eqref{eq:fullhamiltonian}. Although we concentrate on a 1D geometry with nearest neighbor interactions, the same model can be readily generalized to higher dimensions and more neighbors by including more frequency components to the driving fields.

Equations~\eqref{eq:PXP1} and \eqref{eq:PXP2} describe an effective PXP model, previously applied to theoretically describe the Rydberg blockade and facilitation constraints in atomic chains~\cite{Lesanovsky2011,Olmos2012facil,Roghani2018,Turner2018,Lesanovsky2019,ostmann2019}, but generalized here to a wider set of unitary and dissipative conditional operators stemming from the multifrequency driving fields. Fig. \ref{fig:QCA_setup}(b) depicts the unitary and non-unitary conditional update rules for the central site $j$ of a three-site neighborhood. Each field component $\theta^k_j$ effectuates transitions when there is precisely $k$ Rydberg excitations in the neighborhood, constrained by the projection operators $\mathrm{P}_{j-1}^{\alpha},\mathrm{P}_{j+1}^{\beta}$ (i.e. $\alpha=\beta=1$ means the state will only change if both left and right neighbors are in state $|1\rangle$). The special case $\theta^{0}_j\neq 0, \theta^{k>0}_j=0$ corresponds to the Rydberg blockade scenario, while for $\theta^{k>0}_j\neq 0, \theta^{0}_j=0$ corresponds to facilitated excitation in the presence of $k$ already excited neighbors. The inclusion of strong dissipative couplings via the second multifrequency field $\phi^k$ realizes an additional set of irreversible conditional interactions that bring atoms back to the $\ket{0}$ state.

\begin{figure}[t]
	\includegraphics[width=1\columnwidth]{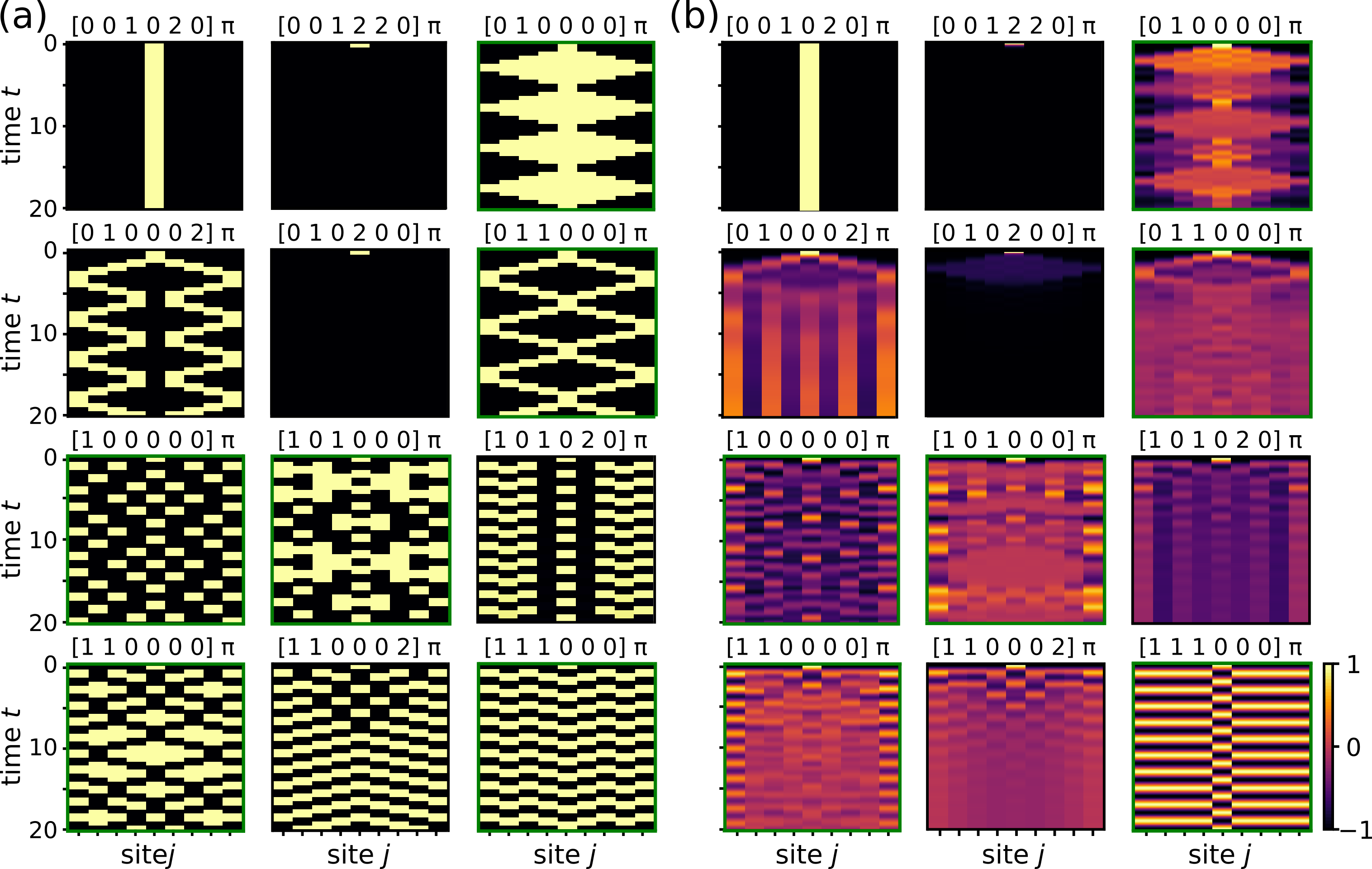}
	\caption{Numerical simulations of quantum dynamics for discrete and continuous time evolution according to different QCA rules starting from the state $\ket{000010000}$. Purely unitary rules are indicated with a green border. (a) Discrete time evolution of the magnetization $\langle \hat Z_j \rangle$ of $N=9$ sites with block partitioning \small{$ABABABABA$} according to Eq.~\eqref{eq:BQCA} for $t=20$ update steps. The different panels correspond to a subset of rules indexed by the parameters $[\theta^0,\theta^1,\theta^2,\tilde \phi^0,\tilde \phi^1, \tilde \phi^2]$. (b) Corresponding continuous time evolution without block partitioning.}
	\label{fig:discrete_continuous}
\end{figure}

\noindent\emph{Numerical simulation of QCA dynamics:-}
The effective two-level representation given by equations \eqref{eq:PXP1} and \eqref{eq:PXP2} can be interpreted as a set of unitary and non-unitary elementary QCA~\cite{Brennen2003}, parameterized by $[\theta^0,\theta^1,\theta^2,\tilde \phi^0,\tilde \phi^1, \tilde \phi^2]$, analogous to the binary string representation used in classical CA. In the following, we consider either discrete or continuous time evolution, described by the application of an (in general non-unitary) operator 
\begin{equation}\label{eq:BQCA}
\rho(t)=(\hat M)^t\rho(0),
\end{equation}
where $\hat  M = \mathrm{exp}({\mathcal{L}})$. For continuous time evolution we treat $t$ as a continuous variable. In the discrete case, a block partitioning scheme is used~\cite{Brennen2003}, meaning that $\hat M = \mathrm{exp}({\mathcal{L}_B})\mathrm{exp}({\mathcal{L}_A})$ is separated according to two sublattices $A$ (odd sites) and $B$ (even sites), which are each updated in alternating fashion an integer number of times $t$. Block partitioned QCA could be experimentally implemented using two different sets of atomic states/species or spatially structuring the Rydberg excitation lasers to address even and odd sites independently. In both cases we solve the master equation using a linear multistep method and the QuTiP package~\cite{qutip2013}.

Fig.~\ref{fig:discrete_continuous} shows numerical simulations of the effective master equation for both discrete time (block partitioned, where we restrict unitary rotations to 0 or $\pi$ and dissipative jump probabilities to $0$ or $1-e^{-2\pi}$ in each step) and the corresponding continuous (non-partitioned) time evolution. We choose 12 representative rule sets (out of $2^6=64$ digital combinations of the parameters $\theta^k,\tilde \phi^k$), which are assumed to be equal for the $A$ and $B$ sublattices. The panels with solid green borders correspond to purely unitary rules ($\tilde \phi^k=0$).  The numerical simulations are performed for 9 atoms, starting from the initial state with the central atom in the $\ket 1$ state and all others in $\ket 0$. This state is evolved for $20$ time units via Eq.~\eqref{eq:BQCA} assuming open boundary conditions (which can be treated as two additional fictitious spins on the left and right fixed to $\ket 0 $).  Bright (dark) colors reflect high (low) magnetization  $\langle \hat Z_j \rangle \approx 1 $ ($\langle \hat Z_j \rangle \approx -1 $). 

The simulated dynamics reveal a variety of different dynamical structures reminiscent of classical CA, including fixed point, periodic, and complex/fractal like structures~(comparable to those studied in Ref.~\cite{hillberry2016}). Furthermore, discrete and continuous time evolution show qualitatively similar features (especially for early times, i.e. time index roughly equal to lattice size), except for a generally lower contrast for the continuous time case. This does not necessarily indicate a loss of coherence however, as it is also seen for the purely unitary rules which can be explained by the build up of entanglement during QCA evolution. 

Careful inspection of the continuous time evolution shows additional periodicities and non-trivial stationary states that are not present in the discrete time evolution. As a specific example we highlight the non-unitary rule $[0,1,0,0,0,2]\times \pi$ (Fig.~\ref{fig:discrete_continuous}-second row, first column). Initially, both discrete and continuous time evolution show similar light-cone like propagation of the excitation. Upon reaching the boundary however, the two cases deviate strongly. Rather than simply reflecting from the boundary, the continuous time evolution evolves toward a steady state that exhibits an antiferromagnetically ordered pattern. Qualitatively this can be understood as the competition between the conditional $k=1$ neighbor driving which favors spreading of the excitations while the $k=2$ neighbor depumping suppresses nearest-neighbor excitations. Thus the final state $\ket{101010101}$ is a dark state for both terms. This highlights the possibility to use unitary and non-unitary QCA dynamics to generate correlated many-body states as the stationary state of the open system dynamics and it is an interesting question whether it can also be used to generate highly entangled quantum states.

\noindent\emph{Steering QCA evolution to highly entangled states:-} Quantum state engineering via open system dynamics is typically cast in terms of finding a Liouvillian $\mathcal L$ that yields a desirable (e.g., entangled) state as the stationary state of the dynamics~\cite{verstraete2009,weimer2010,rao2013,Carr2013,rao2014,Shao2014,Roghani2018,Ding2019}. However in general it is a hard problem to find $\mathcal L$ (and a corresponding set of physically available interactions) that result in this state. We show here that an appropriate combination of QCA rules may be found that steer quantum dynamics into desired quantum states on demand.

\begin{figure}[t]
	\centering
	\includegraphics[width=1\columnwidth]{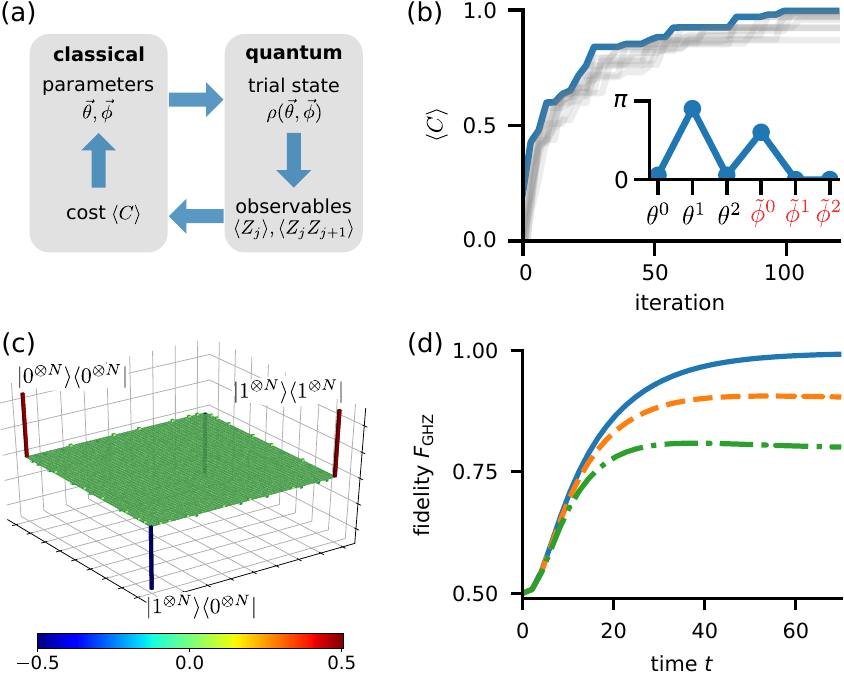}
	\caption{Variational quantum optimization of the Rydberg QCA stationary state towards highly entangled states for a chain of $N=6$ sites. (a) Hybrid quantum-classical feedback loop used to steer the QCA dynamics to desired quantum states. (b) Convergence of particle swarm optimization with a population size of 10 (gray lines) towards states with large covariance coefficient $\braket C$. The solid blue line highlights the best individual. The optimal parameters are shown in the inset. (c) Graphical representation of the density matrix of the optimized state (imaginary parts are $<10^{-3}$). (d) Time evolution of the fidelity between the resulting QCA state using the optimal variational parameters and the GHZ$^N$ state. The system evolves to a highly entangled state within approximately 70 time units. The dashed orange (dashed-dotted green) line shows the same evolution including a Rydberg state decay rate of $\gamma/2\pi = 0.8\,$kHz  ($2.4\,$kHz) (see the text for other parameters).} 
	\label{fig:VSO}
\end{figure}
The basic idea is to embed the Rydberg QCA within a variational optimization loop which iteratively adjusts the QCA parameters to reach a desired target state as outlined in Fig.~\ref{fig:VSO}(a). The role of the quantum system is to generate trial states according to these parameters and to allow measurements of suitable observables that reflect the desired (quantum) correlations. The outcome of these measurements is used to compute a classical cost function which is then minimized or maximized by a classical search algorithm. While this variational approach has been powerfully demonstrated for finding ground states of unitary quantum systems \cite{peruzzo2014variational, omalley2016scalable, kandala2017hardware, kokail2019self}, we show it can also apply to non-unitary quantum evolution and stationary states.

To steer the system to highly correlated states we choose to maximize elements of the covariance matrix $C_{i,j} = \mathrm{Tr}[\rho (\hat Z_i-\braket{\hat Z_i})(\hat Z_j-\braket{\hat Z_j})]$. For the following we average over all neighbours, i.e. $\braket C = 1/N \sum_j C_{j,j+1} = 0$ for a separable state while $\braket C=\pm 1$ for a pairwise inseparable state. Here we restrict to proof-of-principle numerical simulations for relatively small system sizes of $N=6$ sites, although this provides valuable guidance for finding optimal parameter regimes for larger systems. In the following we numerically solve for the steady state of the continuous time QCA evolution with periodic boundary conditions (to minimize edge effects for the considered system sizes) and global variational parameters $\theta^k,\tilde \phi^k$. We use $\ket{0}^{\otimes N}=\ket{000000}$ as the initial state, but we observe similar behavior for other initial states. We maximize $\braket C$ using the particle swarm optimization (PSO) algorithm~\cite{kennedy2010particle}. 

Figure~\ref{fig:VSO}(b) shows the convergence of the variational optimization algorithm as a function of the number of PSO iterations. We use a population of 10 individuals (gray lines), with the best individual highlighted in blue. We find that convergence is robust and relatively fast (within 100 iterations), saturating at a value close to the maximum value $\braket C=1$. Inspecting the resulting density matrix in Fig.~\ref{fig:VSO}(c), we find that the final state is very close to the $\ket{\mathrm{GHZ}^N}= 1/\sqrt{2} (\ket{0}^{\otimes N}-\ket{1}^{\otimes N})$ state. This can be understood since both $\ket{0}^{\otimes N}$ and $\ket{1}^{\otimes N}$ are dark with respect to the projectors associated to $\theta^1$ and $ \tilde{\phi^0}$. However, as these separable states both yield $\braket{C}=0$, it appears that a relatively weak contribution from $\theta^0$ and $\theta^2$ is important to stabilize the $\ket{\mathrm{GHZ}^N}$ state with a well defined relative phase.

The evolution towards the $\ket{\mathrm{GHZ}^N}$ state using the optimized parameters is shown in Fig.~\ref{fig:VSO}(d), quantified by the fidelity $F_\text{GHZ}(t) = \text{Tr}\left [ \sqrt{\sqrt{\rho(t)} \ket{\mathrm{GHZ}^N}\bra{\mathrm{GHZ}^N} \sqrt{\rho(t)} }\right ]^2$, which reaches $\gtrsim 0.99$ within $70$ time units. This is slower than a comparable protocol using purely unitary (discrete time, block-partitioned) evolution by repetitively applying the rule $[0,\pi,0,0,0,0]$ starting with the central qubit in $(\ket{0}+\ket{1})/\sqrt{2}$, which requires only $(N-1)/2$ time steps~\cite{Brennen2003} (see also Ref.~\cite{Omran570} for a related experimental protocol). However the dissipative protocol has the advantage that it is not sensitive to the precise timings or the initial state. The GHZ$^N$ state is very promising as a resource for quantum metrology~\cite{dur2014improved} and measurement-based quantum computing~\cite{nickerson2013topological}, as it is, e.g., a stabilizer state (+1 co-eigenstate) of stabilizer operators generated from the set of $n$ independent operators $\{Z_1Z_2, Z_2Z_3,...,Z_{n-1}Z_n, X_1X_2...X_n \}$. We have also performed minimization of $\braket{C}$ and the resulting solution is the antiferromagnetic GHZ state $\ket{\mathrm{AF}}= 1/\sqrt{2} (\ket{010101}-\ket{101010})$ (see Supplemental Material for more details), which shows that variational quantum optimization combined with Rydberg QCA provides a powerful and rather general approach to quantum state engineering.

\noindent\emph{Experimental feasibility:-} Rydberg QCA could be implemented in a number of physical systems that support a simple (three) level scheme and strong state-dependent nearest-neighbor interactions. But to estimate realistic experimental parameters we consider an array of ultracold $^{39}$K atoms. To minimize the effects of atomic motion and position fluctuations one could make use of blue-detuned lasers to simultaneously trap both ground and Rydberg states~\cite{barredo2019three, graham2019rydberg} and cool the atoms close to the ground state in each trap~\cite{Kaufman2012}. The $\ket{g}\leftrightarrow\ket{r}$ coupling could be achieved using a two-photon resonance with large detuning from the $\ket{e}$ state, while $\ket{r}\leftrightarrow\ket{e}$ could be a single-photon transition, with multiple tones generated by electro-optical modulators. Parameters corresponding to those used for Fig.~\ref{fig:discrete_continuous} (and Fig.~\ref{fig:VSO}) are $V/2\pi = 50\,$MHz (typical for the $\ket{r}=\ket{80s_{1/2}}$ state at a distance of $a=6\,\mu$m), $\theta^k/2\pi \leq 1\,$MHz, $\tilde \phi^k/2\pi \leq 2\,$MHz, which gives a characteristic time unit of $t=\pi/\theta^k= 500\,$ns. Therefore, within a typical Rydberg state lifetime ($\gamma^{-1} \approx 200\,\mu$s) one could realize up to $\sim 400$ time steps and prepare highly entangled states with high fidelity. To quantify the effect of uncontrolled dissipative processes, we also perform simulations including jump operators describing decay of the Rydberg state with a rate $\gamma$. Figure~\ref{fig:VSO}(d) shows the fidelity for preparing a 6 atom GHZ$^N$ state still reaches $F\gtrsim 0.9$ for $\gamma/2\pi=0.8\,$kHz, and  $F\gtrsim 0.8$ when accounting for additional losses ($\gamma/2\pi=2.4\,$kHz) which could arise due to off-resonant coupling to other short-lived intermediate states encountered in experiments~\cite{Omran570}. The fidelity could be further improved using error correction schemes~\cite{fitzsimons2009quantum, paz2011bulk} or using longer-lived Rydberg states in cryogenic environments~\cite{nguyen2018towards}. 

To conclude, we have put forward a promising approach to quantum state engineering and QIP that is highly parallelizable and exploits both unitary and non-unitary multiqubit interactions. Already a very basic set of conditional QCA rules acting under continuous time evolution can generate a rich variety of complex quantum dynamics and highly entangled states. Allowing for different rules to be applied at different times would enable the generation of deeper quantum circuits opening up the possibility for universal quantum computing~\cite{lloyd1993,raussendorf2005,vollbrecht2006reversible, watrous1995one}. 

\acknowledgements{We acknowledge discussions with Pierre Collet, Guido Pupillo, and Matthias Weidem\"uller.  This work is supported by the `Investissements d'Avenir' programme through the Excellence Initiative of the University of Strasbourg (IdEx), the University of Strasbourg Institute for Advanced Study (USIAS) and is part of and supported by the DFG Collaborative Research Center `SFB 1225 (ISOQUANT)'. G.K.B. receives support from the Australian Research Council Center of Excellence for Engineered Quantum Systems (Project number CE170100009). S.S. and T.M.W. acknowledge support from the French National Research Agency (ANR) through the Programme d'Investissement d'Avenir under contract ANR-17-EURE-0024.}


\bibliography{rydbergQCA}

\begin{thebibliography}{68}%
\makeatletter
\providecommand \@ifxundefined [1]{%
 \@ifx{#1\undefined}
}%
\providecommand \@ifnum [1]{%
 \ifnum #1\expandafter \@firstoftwo
 \else \expandafter \@secondoftwo
 \fi
}%
\providecommand \@ifx [1]{%
 \ifx #1\expandafter \@firstoftwo
 \else \expandafter \@secondoftwo
 \fi
}%
\providecommand \natexlab [1]{#1}%
\providecommand \enquote  [1]{``#1''}%
\providecommand \bibnamefont  [1]{#1}%
\providecommand \bibfnamefont [1]{#1}%
\providecommand \citenamefont [1]{#1}%
\providecommand \href@noop [0]{\@secondoftwo}%
\providecommand \href [0]{\begingroup \@sanitize@url \@href}%
\providecommand \@href[1]{\@@startlink{#1}\@@href}%
\providecommand \@@href[1]{\endgroup#1\@@endlink}%
\providecommand \@sanitize@url [0]{\catcode `\\12\catcode `\$12\catcode
  `\&12\catcode `\#12\catcode `\^12\catcode `\_12\catcode `\%12\relax}%
\providecommand \@@startlink[1]{}%
\providecommand \@@endlink[0]{}%
\providecommand \url  [0]{\begingroup\@sanitize@url \@url }%
\providecommand \@url [1]{\endgroup\@href {#1}{\urlprefix }}%
\providecommand \urlprefix  [0]{URL }%
\providecommand \Eprint [0]{\href }%
\providecommand \doibase [0]{http://dx.doi.org/}%
\providecommand \selectlanguage [0]{\@gobble}%
\providecommand \bibinfo  [0]{\@secondoftwo}%
\providecommand \bibfield  [0]{\@secondoftwo}%
\providecommand \translation [1]{[#1]}%
\providecommand \BibitemOpen [0]{}%
\providecommand \bibitemStop [0]{}%
\providecommand \bibitemNoStop [0]{.\EOS\space}%
\providecommand \EOS [0]{\spacefactor3000\relax}%
\providecommand \BibitemShut  [1]{\csname bibitem#1\endcsname}%
\let\auto@bib@innerbib\@empty
\bibitem [{\citenamefont {Preskill}(2018)}]{Preskill2018}%
  \BibitemOpen
  \bibfield  {author} {\bibinfo {author} {\bibfnamefont {J.}~\bibnamefont
  {Preskill}},\ }\bibfield  {title} {\enquote {\bibinfo {title} {Quantum
  computing in the {NISQ} era and beyond},}\ }\href {\doibase
  10.22331/q-2018-08-06-79} {\bibfield  {journal} {\bibinfo  {journal}
  {Quantum}\ }\textbf {\bibinfo {volume} {2}},\ \bibinfo {pages} {79} (\bibinfo
  {year} {2018})}\BibitemShut {NoStop}%
\bibitem [{\citenamefont {Lukin}\ \emph {et~al.}(2001)\citenamefont {Lukin},
  \citenamefont {Fleischhauer}, \citenamefont {Cote}, \citenamefont {Duan},
  \citenamefont {Jaksch}, \citenamefont {Cirac},\ and\ \citenamefont
  {Zoller}}]{Lukin2001}%
  \BibitemOpen
  \bibfield  {author} {\bibinfo {author} {\bibfnamefont {M.~D.}\ \bibnamefont
  {Lukin}}, \bibinfo {author} {\bibfnamefont {M.}~\bibnamefont {Fleischhauer}},
  \bibinfo {author} {\bibfnamefont {R.}~\bibnamefont {Cote}}, \bibinfo {author}
  {\bibfnamefont {L.~M.}\ \bibnamefont {Duan}}, \bibinfo {author}
  {\bibfnamefont {D.}~\bibnamefont {Jaksch}}, \bibinfo {author} {\bibfnamefont
  {J.~I.}\ \bibnamefont {Cirac}}, \ and\ \bibinfo {author} {\bibfnamefont
  {P.}~\bibnamefont {Zoller}},\ }\bibfield  {title} {\enquote {\bibinfo {title}
  {Dipole blockade and quantum information processing in mesoscopic atomic
  ensembles},}\ }\href {\doibase 10.1103/PhysRevLett.87.037901} {\bibfield
  {journal} {\bibinfo  {journal} {Phys. Rev. Lett.}\ }\textbf {\bibinfo
  {volume} {87}},\ \bibinfo {pages} {037901} (\bibinfo {year}
  {2001})}\BibitemShut {NoStop}%
\bibitem [{\citenamefont {Saffman}\ \emph {et~al.}(2010)\citenamefont
  {Saffman}, \citenamefont {Walker},\ and\ \citenamefont
  {M{\o}lmer}}]{saffman2010}%
  \BibitemOpen
  \bibfield  {author} {\bibinfo {author} {\bibfnamefont {M.}~\bibnamefont
  {Saffman}}, \bibinfo {author} {\bibfnamefont {T.~G.}\ \bibnamefont {Walker}},
  \ and\ \bibinfo {author} {\bibfnamefont {K.}~\bibnamefont {M{\o}lmer}},\
  }\bibfield  {title} {\enquote {\bibinfo {title} {Quantum information with
  {R}ydberg atoms},}\ }\href {\doibase 10.1103/RevModPhys.82.2313} {\bibfield
  {journal} {\bibinfo  {journal} {Rev. Mod. Phys}\ }\textbf {\bibinfo {volume}
  {82}},\ \bibinfo {pages} {2313} (\bibinfo {year} {2010})}\BibitemShut
  {NoStop}%
\bibitem [{\citenamefont {Weimer}\ \emph {et~al.}(2010)\citenamefont {Weimer},
  \citenamefont {M{\"u}ller}, \citenamefont {Lesanovsky}, \citenamefont
  {Zoller},\ and\ \citenamefont {B{\"u}chler}}]{weimer2010}%
  \BibitemOpen
  \bibfield  {author} {\bibinfo {author} {\bibfnamefont {H.}~\bibnamefont
  {Weimer}}, \bibinfo {author} {\bibfnamefont {M.}~\bibnamefont {M{\"u}ller}},
  \bibinfo {author} {\bibfnamefont {I.}~\bibnamefont {Lesanovsky}}, \bibinfo
  {author} {\bibfnamefont {P.}~\bibnamefont {Zoller}}, \ and\ \bibinfo {author}
  {\bibfnamefont {H.~P.}\ \bibnamefont {B{\"u}chler}},\ }\bibfield  {title}
  {\enquote {\bibinfo {title} {A {R}ydberg quantum simulator},}\ }\href
  {\doibase 10.1038/nphys1614} {\bibfield  {journal} {\bibinfo  {journal} {Nat.
  Phys.}\ }\textbf {\bibinfo {volume} {6}},\ \bibinfo {pages} {382} (\bibinfo
  {year} {2010})}\BibitemShut {NoStop}%
\bibitem [{\citenamefont {Saffman}(2016)}]{saffman2016}%
  \BibitemOpen
  \bibfield  {author} {\bibinfo {author} {\bibfnamefont {M.}~\bibnamefont
  {Saffman}},\ }\bibfield  {title} {\enquote {\bibinfo {title} {Quantum
  computing with atomic qubits and {R}ydberg interactions: progress and
  challenges},}\ }\href {\doibase 10.1088/0953-4075/49/20/202001} {\bibfield
  {journal} {\bibinfo  {journal} {J. Phys. B: At. Mol. Opt. Phys}\ }\textbf
  {\bibinfo {volume} {49}},\ \bibinfo {pages} {202001} (\bibinfo {year}
  {2016})}\BibitemShut {NoStop}%
\bibitem [{\citenamefont {Unanyan}\ and\ \citenamefont
  {Fleischhauer}(2002)}]{Unanyan2002}%
  \BibitemOpen
  \bibfield  {author} {\bibinfo {author} {\bibfnamefont {R.~G.}\ \bibnamefont
  {Unanyan}}\ and\ \bibinfo {author} {\bibfnamefont {M.}~\bibnamefont
  {Fleischhauer}},\ }\bibfield  {title} {\enquote {\bibinfo {title} {Efficient
  and robust entanglement generation in a many-particle system with resonant
  dipole-dipole interactions},}\ }\href {\doibase 10.1103/PhysRevA.66.032109}
  {\bibfield  {journal} {\bibinfo  {journal} {Phys. Rev. A}\ }\textbf {\bibinfo
  {volume} {66}},\ \bibinfo {pages} {032109} (\bibinfo {year}
  {2002})}\BibitemShut {NoStop}%
\bibitem [{\citenamefont {Isenhower}\ \emph {et~al.}(2011)\citenamefont
  {Isenhower}, \citenamefont {Saffman},\ and\ \citenamefont
  {M{\o}lmer}}]{Isenhower2011}%
  \BibitemOpen
  \bibfield  {author} {\bibinfo {author} {\bibfnamefont {L.}~\bibnamefont
  {Isenhower}}, \bibinfo {author} {\bibfnamefont {M.}~\bibnamefont {Saffman}},
  \ and\ \bibinfo {author} {\bibfnamefont {K.}~\bibnamefont {M{\o}lmer}},\
  }\bibfield  {title} {\enquote {\bibinfo {title} {Multibit {C}k{NOT} quantum
  gates via {R}ydberg blockade},}\ }\href {\doibase 10.1007/s11128-011-0292-4}
  {\bibfield  {journal} {\bibinfo  {journal} {Quantum Inf. Process.}\ }\textbf
  {\bibinfo {volume} {10}},\ \bibinfo {pages} {755} (\bibinfo {year}
  {2011})}\BibitemShut {NoStop}%
\bibitem [{\citenamefont {Ebert}\ \emph {et~al.}(2015)\citenamefont {Ebert},
  \citenamefont {Kwon}, \citenamefont {Walker},\ and\ \citenamefont
  {Saffman}}]{Ebert2015}%
  \BibitemOpen
  \bibfield  {author} {\bibinfo {author} {\bibfnamefont {M.}~\bibnamefont
  {Ebert}}, \bibinfo {author} {\bibfnamefont {M.}~\bibnamefont {Kwon}},
  \bibinfo {author} {\bibfnamefont {T.~G.}\ \bibnamefont {Walker}}, \ and\
  \bibinfo {author} {\bibfnamefont {M.}~\bibnamefont {Saffman}},\ }\bibfield
  {title} {\enquote {\bibinfo {title} {Coherence and {R}ydberg blockade of
  atomic ensemble qubits},}\ }\href {\doibase 10.1103/PhysRevLett.115.093601}
  {\bibfield  {journal} {\bibinfo  {journal} {Phys. Rev. Lett.}\ }\textbf
  {\bibinfo {volume} {115}},\ \bibinfo {pages} {093601} (\bibinfo {year}
  {2015})}\BibitemShut {NoStop}%
\bibitem [{\citenamefont {Zeiher}\ \emph {et~al.}(2015)\citenamefont {Zeiher},
  \citenamefont {Schau\ss{}}, \citenamefont {Hild}, \citenamefont {Macr\`{\i}},
  \citenamefont {Bloch},\ and\ \citenamefont {Gross}}]{Zeiher2015}%
  \BibitemOpen
  \bibfield  {author} {\bibinfo {author} {\bibfnamefont {J.}~\bibnamefont
  {Zeiher}}, \bibinfo {author} {\bibfnamefont {P.}~\bibnamefont {Schau\ss{}}},
  \bibinfo {author} {\bibfnamefont {S.}~\bibnamefont {Hild}}, \bibinfo {author}
  {\bibfnamefont {T.}~\bibnamefont {Macr\`{\i}}}, \bibinfo {author}
  {\bibfnamefont {I.}~\bibnamefont {Bloch}}, \ and\ \bibinfo {author}
  {\bibfnamefont {C.}~\bibnamefont {Gross}},\ }\bibfield  {title} {\enquote
  {\bibinfo {title} {Microscopic characterization of scalable coherent
  {R}ydberg superatoms},}\ }\href {\doibase 10.1103/PhysRevX.5.031015}
  {\bibfield  {journal} {\bibinfo  {journal} {Phys. Rev. X}\ }\textbf {\bibinfo
  {volume} {5}},\ \bibinfo {pages} {031015} (\bibinfo {year}
  {2015})}\BibitemShut {NoStop}%
\bibitem [{\citenamefont {Gambetta}\ \emph {et~al.}(2019)\citenamefont
  {Gambetta}, \citenamefont {Li}, \citenamefont {Schmidt-Kaler},\ and\
  \citenamefont {Lesanovsky}}]{gambetta2019}%
  \BibitemOpen
  \bibfield  {author} {\bibinfo {author} {\bibfnamefont {F.~M.}\ \bibnamefont
  {Gambetta}}, \bibinfo {author} {\bibfnamefont {W.}~\bibnamefont {Li}},
  \bibinfo {author} {\bibfnamefont {F.}~\bibnamefont {Schmidt-Kaler}}, \ and\
  \bibinfo {author} {\bibfnamefont {I.}~\bibnamefont {Lesanovsky}},\ }\bibfield
   {title} {\enquote {\bibinfo {title} {Engineering non-binary {R}ydberg
  interactions via electron-phonon coupling},}\ }\href
  {https://arxiv.org/abs/1907.11664} {\bibfield  {journal} {\bibinfo  {journal}
  {arXiv:1907.11664}\ } (\bibinfo {year} {2019})}\BibitemShut {NoStop}%
\bibitem [{\citenamefont {Verstraete}\ \emph {et~al.}(2009)\citenamefont
  {Verstraete}, \citenamefont {Wolf},\ and\ \citenamefont
  {Cirac}}]{verstraete2009}%
  \BibitemOpen
  \bibfield  {author} {\bibinfo {author} {\bibfnamefont {F.}~\bibnamefont
  {Verstraete}}, \bibinfo {author} {\bibfnamefont {M.~M.}\ \bibnamefont
  {Wolf}}, \ and\ \bibinfo {author} {\bibfnamefont {J.~I.}\ \bibnamefont
  {Cirac}},\ }\bibfield  {title} {\enquote {\bibinfo {title} {Quantum
  computation and quantum-state engineering driven by dissipation},}\ }\href
  {\doibase 10.1038/nphys1342} {\bibfield  {journal} {\bibinfo  {journal} {Nat.
  Phys.}\ }\textbf {\bibinfo {volume} {5}},\ \bibinfo {pages} {633} (\bibinfo
  {year} {2009})}\BibitemShut {NoStop}%
\bibitem [{\citenamefont {Rao}\ and\ \citenamefont
  {M{\o}lmer}(2013)}]{rao2013}%
  \BibitemOpen
  \bibfield  {author} {\bibinfo {author} {\bibfnamefont {D.~D.~B.}\
  \bibnamefont {Rao}}\ and\ \bibinfo {author} {\bibfnamefont {K.}~\bibnamefont
  {M{\o}lmer}},\ }\bibfield  {title} {\enquote {\bibinfo {title} {Dark
  entangled steady states of interacting {R}ydberg atoms},}\ }\href {\doibase
  10.1103/PhysRevLett.111.033606} {\bibfield  {journal} {\bibinfo  {journal}
  {Phys. Rev. Lett.}\ }\textbf {\bibinfo {volume} {111}},\ \bibinfo {pages}
  {033606} (\bibinfo {year} {2013})}\BibitemShut {NoStop}%
\bibitem [{\citenamefont {Carr}\ and\ \citenamefont
  {Saffman}(2013)}]{Carr2013}%
  \BibitemOpen
  \bibfield  {author} {\bibinfo {author} {\bibfnamefont {A.~W.}\ \bibnamefont
  {Carr}}\ and\ \bibinfo {author} {\bibfnamefont {M.}~\bibnamefont {Saffman}},\
  }\bibfield  {title} {\enquote {\bibinfo {title} {Preparation of entangled and
  antiferromagnetic states by dissipative {R}ydberg pumping},}\ }\href
  {\doibase 10.1103/PhysRevLett.111.033607} {\bibfield  {journal} {\bibinfo
  {journal} {Phys. Rev. Lett.}\ }\textbf {\bibinfo {volume} {111}},\ \bibinfo
  {pages} {033607} (\bibinfo {year} {2013})}\BibitemShut {NoStop}%
\bibitem [{\citenamefont {Rao}\ and\ \citenamefont
  {M{\o}lmer}(2014)}]{rao2014}%
  \BibitemOpen
  \bibfield  {author} {\bibinfo {author} {\bibfnamefont {D.~D.~B.}\
  \bibnamefont {Rao}}\ and\ \bibinfo {author} {\bibfnamefont {K.}~\bibnamefont
  {M{\o}lmer}},\ }\bibfield  {title} {\enquote {\bibinfo {title} {Deterministic
  entanglement of {R}ydberg ensembles by engineered dissipation},}\ }\href
  {\doibase 10.1103/PhysRevA.90.062319} {\bibfield  {journal} {\bibinfo
  {journal} {Phys. Rev. A}\ }\textbf {\bibinfo {volume} {90}},\ \bibinfo
  {pages} {062319} (\bibinfo {year} {2014})}\BibitemShut {NoStop}%
\bibitem [{\citenamefont {Shao}\ \emph {et~al.}(2014)\citenamefont {Shao},
  \citenamefont {You}, \citenamefont {Zheng}, \citenamefont {Oh},\ and\
  \citenamefont {Zhang}}]{Shao2014}%
  \BibitemOpen
  \bibfield  {author} {\bibinfo {author} {\bibfnamefont {X.~Q.}\ \bibnamefont
  {Shao}}, \bibinfo {author} {\bibfnamefont {J.~B.}\ \bibnamefont {You}},
  \bibinfo {author} {\bibfnamefont {T.~Y.}\ \bibnamefont {Zheng}}, \bibinfo
  {author} {\bibfnamefont {C.~H.}\ \bibnamefont {Oh}}, \ and\ \bibinfo {author}
  {\bibfnamefont {S.}~\bibnamefont {Zhang}},\ }\bibfield  {title} {\enquote
  {\bibinfo {title} {Stationary three-dimensional entanglement via dissipative
  {R}ydberg pumping},}\ }\href {\doibase 10.1103/PhysRevA.89.052313} {\bibfield
   {journal} {\bibinfo  {journal} {Phys. Rev. A}\ }\textbf {\bibinfo {volume}
  {89}},\ \bibinfo {pages} {052313} (\bibinfo {year} {2014})}\BibitemShut
  {NoStop}%
\bibitem [{\citenamefont {Roghani}\ and\ \citenamefont
  {Weimer}(2018)}]{Roghani2018}%
  \BibitemOpen
  \bibfield  {author} {\bibinfo {author} {\bibfnamefont {M.}~\bibnamefont
  {Roghani}}\ and\ \bibinfo {author} {\bibfnamefont {H.}~\bibnamefont
  {Weimer}},\ }\bibfield  {title} {\enquote {\bibinfo {title} {Dissipative
  preparation of entangled many-body states with {R}ydberg atoms},}\ }\href
  {\doibase 10.1088/2058-9565/aab3f3} {\bibfield  {journal} {\bibinfo
  {journal} {Quantum Sci. Technol.}\ }\textbf {\bibinfo {volume} {3}},\
  \bibinfo {pages} {035002} (\bibinfo {year} {2018})}\BibitemShut {NoStop}%
\bibitem [{\citenamefont {Ding}\ \emph {et~al.}(2019)\citenamefont {Ding},
  \citenamefont {Hu}, \citenamefont {Shen}, \citenamefont {Yang}, \citenamefont
  {Wu},\ and\ \citenamefont {Zheng}}]{Ding2019}%
  \BibitemOpen
  \bibfield  {author} {\bibinfo {author} {\bibfnamefont {Z.~X.}\ \bibnamefont
  {Ding}}, \bibinfo {author} {\bibfnamefont {C.~S.}\ \bibnamefont {Hu}},
  \bibinfo {author} {\bibfnamefont {L.~T.}\ \bibnamefont {Shen}}, \bibinfo
  {author} {\bibfnamefont {Z.~B.}\ \bibnamefont {Yang}}, \bibinfo {author}
  {\bibfnamefont {H.}~\bibnamefont {Wu}}, \ and\ \bibinfo {author}
  {\bibfnamefont {S.~B.}\ \bibnamefont {Zheng}},\ }\bibfield  {title} {\enquote
  {\bibinfo {title} {Dissipative entanglement preparation via {R}ydberg
  antiblockade and {L}yapunov control},}\ }\href {\doibase
  10.1088/1612-202x/ab0c8b} {\bibfield  {journal} {\bibinfo  {journal} {Laser
  Phys. Lett.}\ }\textbf {\bibinfo {volume} {16}},\ \bibinfo {pages} {045203}
  (\bibinfo {year} {2019})}\BibitemShut {NoStop}%
\bibitem [{\citenamefont {Wiesner}(2009)}]{Wiesner2009}%
  \BibitemOpen
  \bibfield  {author} {\bibinfo {author} {\bibfnamefont {K.}~\bibnamefont
  {Wiesner}},\ }\enquote {\bibinfo {title} {Quantum cellular automata},}\ in\
  \href {\doibase 10.1007/978-0-387-30440-3_426} {\emph {\bibinfo {booktitle}
  {Encyclopedia of Complexity and Systems Science}}}\ (\bibinfo  {publisher}
  {Springer New York},\ \bibinfo {address} {New York, NY},\ \bibinfo {year}
  {2009})\ pp.\ \bibinfo {pages} {7154--7164}\BibitemShut {NoStop}%
\bibitem [{\citenamefont {Farrelly}(2019)}]{farrelly2019}%
  \BibitemOpen
  \bibfield  {author} {\bibinfo {author} {\bibfnamefont {T.}~\bibnamefont
  {Farrelly}},\ }\bibfield  {title} {\enquote {\bibinfo {title} {A review of
  quantum cellular automata},}\ }\href {https://arxiv.org/abs/1904.13318}
  {\bibfield  {journal} {\bibinfo  {journal} {arXiv:1904.13318}\ } (\bibinfo
  {year} {2019})}\BibitemShut {NoStop}%
\bibitem [{\citenamefont {Arrighi}(2019)}]{arrighi2019}%
  \BibitemOpen
  \bibfield  {author} {\bibinfo {author} {\bibfnamefont {P.}~\bibnamefont
  {Arrighi}},\ }\bibfield  {title} {\enquote {\bibinfo {title} {An overview of
  quantum cellular automata},}\ }\href {\doibase 10.1007/s11047-019-09762-6}
  {\bibfield  {journal} {\bibinfo  {journal} {Nat. Comput.}\ }\textbf {\bibinfo
  {volume} {18}},\ \bibinfo {pages} {885--899} (\bibinfo {year}
  {2019})}\BibitemShut {NoStop}%
\bibitem [{\citenamefont {Lloyd}(1993)}]{lloyd1993}%
  \BibitemOpen
  \bibfield  {author} {\bibinfo {author} {\bibfnamefont {S.}~\bibnamefont
  {Lloyd}},\ }\bibfield  {title} {\enquote {\bibinfo {title} {A potentially
  realizable quantum computer},}\ }\href {\doibase
  10.1126/science.261.5128.1569} {\bibfield  {journal} {\bibinfo  {journal}
  {Science}\ }\textbf {\bibinfo {volume} {261}},\ \bibinfo {pages} {1569--1571}
  (\bibinfo {year} {1993})}\BibitemShut {NoStop}%
\bibitem [{\citenamefont {Watrous}(1995)}]{watrous1995one}%
  \BibitemOpen
  \bibfield  {author} {\bibinfo {author} {\bibfnamefont {J.}~\bibnamefont
  {Watrous}},\ }\bibfield  {title} {\enquote {\bibinfo {title} {On
  one-dimensional quantum cellular automata},}\ }in\ \href {\doibase
  10.1109/SFCS.1995.492583} {\emph {\bibinfo {booktitle} {Proceedings of IEEE
  36th Annual Foundations of Computer Science}}}\ (\bibinfo {year}
  {1995})\BibitemShut {NoStop}%
\bibitem [{\citenamefont {Raussendorf}(2005)}]{raussendorf2005}%
  \BibitemOpen
  \bibfield  {author} {\bibinfo {author} {\bibfnamefont {R.}~\bibnamefont
  {Raussendorf}},\ }\bibfield  {title} {\enquote {\bibinfo {title} {Quantum
  cellular automaton for universal quantum computation},}\ }\href {\doibase
  10.1103/PhysRevA.72.022301} {\bibfield  {journal} {\bibinfo  {journal} {Phys.
  Rev. A}\ }\textbf {\bibinfo {volume} {72}},\ \bibinfo {pages} {022301}
  (\bibinfo {year} {2005})}\BibitemShut {NoStop}%
\bibitem [{\citenamefont {Benjamin}(2000)}]{Benjamin2000}%
  \BibitemOpen
  \bibfield  {author} {\bibinfo {author} {\bibfnamefont {S.~C.}\ \bibnamefont
  {Benjamin}},\ }\bibfield  {title} {\enquote {\bibinfo {title} {Schemes for
  parallel quantum computation without local control of qubits},}\ }\href
  {\doibase 10.1103/PhysRevA.61.020301} {\bibfield  {journal} {\bibinfo
  {journal} {Phys. Rev. A}\ }\textbf {\bibinfo {volume} {61}},\ \bibinfo
  {pages} {020301} (\bibinfo {year} {2000})}\BibitemShut {NoStop}%
\bibitem [{\citenamefont {Benjamin}(2001)}]{Benjamin2001}%
  \BibitemOpen
  \bibfield  {author} {\bibinfo {author} {\bibfnamefont {S.~C.}\ \bibnamefont
  {Benjamin}},\ }\bibfield  {title} {\enquote {\bibinfo {title} {Quantum
  computing without local control of qubit-qubit interactions},}\ }\href
  {\doibase 10.1103/PhysRevLett.88.017904} {\bibfield  {journal} {\bibinfo
  {journal} {Phys. Rev. Lett.}\ }\textbf {\bibinfo {volume} {88}},\ \bibinfo
  {pages} {017904} (\bibinfo {year} {2001})}\BibitemShut {NoStop}%
\bibitem [{\citenamefont {Twamley}(2003)}]{twamley2003}%
  \BibitemOpen
  \bibfield  {author} {\bibinfo {author} {\bibfnamefont {J.}~\bibnamefont
  {Twamley}},\ }\bibfield  {title} {\enquote {\bibinfo {title}
  {Quantum-cellular-automata quantum computing with endohedral fullerenes},}\
  }\href {\doibase 10.1103/PhysRevA.67.052318} {\bibfield  {journal} {\bibinfo
  {journal} {Phys. Rev. A}\ }\textbf {\bibinfo {volume} {67}},\ \bibinfo
  {pages} {052318} (\bibinfo {year} {2003})}\BibitemShut {NoStop}%
\bibitem [{\citenamefont {Brennen}\ and\ \citenamefont
  {Williams}(2003)}]{Brennen2003}%
  \BibitemOpen
  \bibfield  {author} {\bibinfo {author} {\bibfnamefont {G.~K.}\ \bibnamefont
  {Brennen}}\ and\ \bibinfo {author} {\bibfnamefont {J.~E.}\ \bibnamefont
  {Williams}},\ }\bibfield  {title} {\enquote {\bibinfo {title} {Entanglement
  dynamics in one-dimensional quantum cellular automata},}\ }\href {\doibase
  10.1103/PhysRevA.68.042311} {\bibfield  {journal} {\bibinfo  {journal} {Phys.
  Rev. A}\ }\textbf {\bibinfo {volume} {68}},\ \bibinfo {pages} {042311}
  (\bibinfo {year} {2003})}\BibitemShut {NoStop}%
\bibitem [{\citenamefont {Vollbrecht}\ and\ \citenamefont
  {Cirac}(2006)}]{vollbrecht2006reversible}%
  \BibitemOpen
  \bibfield  {author} {\bibinfo {author} {\bibfnamefont {K.~G.~H.}\
  \bibnamefont {Vollbrecht}}\ and\ \bibinfo {author} {\bibfnamefont {J.~I.}\
  \bibnamefont {Cirac}},\ }\bibfield  {title} {\enquote {\bibinfo {title}
  {Reversible universal quantum computation within translation-invariant
  systems},}\ }\href {\doibase 10.1103/PhysRevA.73.012324} {\bibfield
  {journal} {\bibinfo  {journal} {Phys. Rev. A}\ }\textbf {\bibinfo {volume}
  {73}},\ \bibinfo {pages} {012324} (\bibinfo {year} {2006})}\BibitemShut
  {NoStop}%
\bibitem [{\citenamefont {Petrosyan}\ \emph {et~al.}(2016)\citenamefont
  {Petrosyan}, \citenamefont {Saffman},\ and\ \citenamefont
  {M{\o}lmer}}]{Petrosyan_2016}%
  \BibitemOpen
  \bibfield  {author} {\bibinfo {author} {\bibfnamefont {D.}~\bibnamefont
  {Petrosyan}}, \bibinfo {author} {\bibfnamefont {M.}~\bibnamefont {Saffman}},
  \ and\ \bibinfo {author} {\bibfnamefont {K.}~\bibnamefont {M{\o}lmer}},\
  }\bibfield  {title} {\enquote {\bibinfo {title} {{G}rover search algorithm
  with {R}ydberg-blockaded atoms: quantum {Monte Carlo} simulations},}\ }\href
  {\doibase 10.1088/0953-4075/49/9/094004} {\bibfield  {journal} {\bibinfo
  {journal} {J. Phys. B: At. Mol. Opt. Phys.}\ }\textbf {\bibinfo {volume}
  {49}},\ \bibinfo {pages} {094004} (\bibinfo {year} {2016})}\BibitemShut
  {NoStop}%
\bibitem [{\citenamefont {Ostmann}\ \emph {et~al.}(2017)\citenamefont
  {Ostmann}, \citenamefont {Minar}, \citenamefont {Marcuzzi}, \citenamefont
  {Levi},\ and\ \citenamefont {Lesanovsky}}]{ostmann2017}%
  \BibitemOpen
  \bibfield  {author} {\bibinfo {author} {\bibfnamefont {M.}~\bibnamefont
  {Ostmann}}, \bibinfo {author} {\bibfnamefont {J.}~\bibnamefont {Minar}},
  \bibinfo {author} {\bibfnamefont {M.}~\bibnamefont {Marcuzzi}}, \bibinfo
  {author} {\bibfnamefont {E.}~\bibnamefont {Levi}}, \ and\ \bibinfo {author}
  {\bibfnamefont {I.}~\bibnamefont {Lesanovsky}},\ }\bibfield  {title}
  {\enquote {\bibinfo {title} {Non-adiabatic quantum state preparation and
  quantum state transport in chains of {R}ydberg atoms},}\ }\href {\doibase
  10.1088/1367-2630/aa983e} {\bibfield  {journal} {\bibinfo  {journal} {New J.
  Phys.}\ }\textbf {\bibinfo {volume} {19}},\ \bibinfo {pages} {123015}
  (\bibinfo {year} {2017})}\BibitemShut {NoStop}%
\bibitem [{\citenamefont {Keating}\ \emph {et~al.}(2013)\citenamefont
  {Keating}, \citenamefont {Goyal}, \citenamefont {Jau}, \citenamefont
  {Biedermann}, \citenamefont {Landahl},\ and\ \citenamefont
  {Deutsch}}]{Keating2013}%
  \BibitemOpen
  \bibfield  {author} {\bibinfo {author} {\bibfnamefont {T.}~\bibnamefont
  {Keating}}, \bibinfo {author} {\bibfnamefont {K.}~\bibnamefont {Goyal}},
  \bibinfo {author} {\bibfnamefont {Y.~Y.}\ \bibnamefont {Jau}}, \bibinfo
  {author} {\bibfnamefont {G.~W.}\ \bibnamefont {Biedermann}}, \bibinfo
  {author} {\bibfnamefont {A.~J.}\ \bibnamefont {Landahl}}, \ and\ \bibinfo
  {author} {\bibfnamefont {I.~H.}\ \bibnamefont {Deutsch}},\ }\bibfield
  {title} {\enquote {\bibinfo {title} {Adiabatic quantum computation with
  {R}ydberg-dressed atoms},}\ }\href {\doibase 10.1103/PhysRevA.87.052314}
  {\bibfield  {journal} {\bibinfo  {journal} {Phys. Rev. A}\ }\textbf {\bibinfo
  {volume} {87}},\ \bibinfo {pages} {052314} (\bibinfo {year}
  {2013})}\BibitemShut {NoStop}%
\bibitem [{\citenamefont {Lechner}\ \emph {et~al.}(2015)\citenamefont
  {Lechner}, \citenamefont {Hauke},\ and\ \citenamefont
  {Zoller}}]{Lechner2015}%
  \BibitemOpen
  \bibfield  {author} {\bibinfo {author} {\bibfnamefont {W.}~\bibnamefont
  {Lechner}}, \bibinfo {author} {\bibfnamefont {P.}~\bibnamefont {Hauke}}, \
  and\ \bibinfo {author} {\bibfnamefont {P.}~\bibnamefont {Zoller}},\
  }\bibfield  {title} {\enquote {\bibinfo {title} {A quantum annealing
  architecture with all-to-all connectivity from local interactions},}\ }\href
  {\doibase 10.1126/sciadv.1500838} {\bibfield  {journal} {\bibinfo  {journal}
  {Sci. Adv.}\ }\textbf {\bibinfo {volume} {1}},\ \bibinfo {pages} {e1500838}
  (\bibinfo {year} {2015})}\BibitemShut {NoStop}%
\bibitem [{\citenamefont {Glaetzle}\ \emph {et~al.}(2017)\citenamefont
  {Glaetzle}, \citenamefont {van Bijnen}, \citenamefont {Zoller},\ and\
  \citenamefont {Lechner}}]{glaetzle2017}%
  \BibitemOpen
  \bibfield  {author} {\bibinfo {author} {\bibfnamefont {A.~W.}\ \bibnamefont
  {Glaetzle}}, \bibinfo {author} {\bibfnamefont {R.~M.~W.}\ \bibnamefont {van
  Bijnen}}, \bibinfo {author} {\bibfnamefont {P.}~\bibnamefont {Zoller}}, \
  and\ \bibinfo {author} {\bibfnamefont {W.}~\bibnamefont {Lechner}},\
  }\bibfield  {title} {\enquote {\bibinfo {title} {A coherent quantum annealer
  with {R}ydberg atoms},}\ }\href {\doibase 10.1038/ncomms15813} {\bibfield
  {journal} {\bibinfo  {journal} {Nat. Commun.}\ }\textbf {\bibinfo {volume}
  {8}},\ \bibinfo {pages} {15813} (\bibinfo {year} {2017})}\BibitemShut
  {NoStop}%
\bibitem [{\citenamefont {Moll}\ \emph {et~al.}(2018)\citenamefont {Moll},
  \citenamefont {Barkoutsos}, \citenamefont {Bishop}, \citenamefont {Chow},
  \citenamefont {Cross}, \citenamefont {Egger}, \citenamefont {Filipp},
  \citenamefont {Fuhrer}, \citenamefont {Gambetta}, \citenamefont {Ganzhorn},
  \citenamefont {Kandala}, \citenamefont {Mezzacapo}, \citenamefont
  {M{\"u}ller}, \citenamefont {Riess}, \citenamefont {Salis}, \citenamefont
  {Smolin}, \citenamefont {Tavernelli},\ and\ \citenamefont
  {Temme}}]{Moll2018}%
  \BibitemOpen
  \bibfield  {author} {\bibinfo {author} {\bibfnamefont {N.}~\bibnamefont
  {Moll}}, \bibinfo {author} {\bibfnamefont {P.}~\bibnamefont {Barkoutsos}},
  \bibinfo {author} {\bibfnamefont {L.~S.}\ \bibnamefont {Bishop}}, \bibinfo
  {author} {\bibfnamefont {J.~M.}\ \bibnamefont {Chow}}, \bibinfo {author}
  {\bibfnamefont {A.}~\bibnamefont {Cross}}, \bibinfo {author} {\bibfnamefont
  {D.~J.}\ \bibnamefont {Egger}}, \bibinfo {author} {\bibfnamefont
  {S.}~\bibnamefont {Filipp}}, \bibinfo {author} {\bibfnamefont
  {A.}~\bibnamefont {Fuhrer}}, \bibinfo {author} {\bibfnamefont {J.~M.}\
  \bibnamefont {Gambetta}}, \bibinfo {author} {\bibfnamefont {M.}~\bibnamefont
  {Ganzhorn}}, \bibinfo {author} {\bibfnamefont {A.}~\bibnamefont {Kandala}},
  \bibinfo {author} {\bibfnamefont {A.}~\bibnamefont {Mezzacapo}}, \bibinfo
  {author} {\bibfnamefont {P.}~\bibnamefont {M{\"u}ller}}, \bibinfo {author}
  {\bibfnamefont {W.}~\bibnamefont {Riess}}, \bibinfo {author} {\bibfnamefont
  {G.}~\bibnamefont {Salis}}, \bibinfo {author} {\bibfnamefont
  {J.}~\bibnamefont {Smolin}}, \bibinfo {author} {\bibfnamefont
  {I.}~\bibnamefont {Tavernelli}}, \ and\ \bibinfo {author} {\bibfnamefont
  {K.}~\bibnamefont {Temme}},\ }\bibfield  {title} {\enquote {\bibinfo {title}
  {Quantum optimization using variational algorithms on near-term quantum
  devices},}\ }\href {\doibase 10.1088/2058-9565/aab822} {\bibfield  {journal}
  {\bibinfo  {journal} {Quantum Sci. Technol.}\ }\textbf {\bibinfo {volume}
  {3}},\ \bibinfo {pages} {030503} (\bibinfo {year} {2018})}\BibitemShut
  {NoStop}%
\bibitem [{\citenamefont {Kokail}\ \emph
  {et~al.}(2019{\natexlab{a}})\citenamefont {Kokail}, \citenamefont {Maier},
  \citenamefont {van Bijnen}, \citenamefont {Brydges}, \citenamefont {Joshi},
  \citenamefont {Jurcevic}, \citenamefont {Muschik}, \citenamefont {Silvi},
  \citenamefont {Blatt}, \citenamefont {Roos} \emph {et~al.}}]{kokail2019}%
  \BibitemOpen
  \bibfield  {author} {\bibinfo {author} {\bibfnamefont {C.}~\bibnamefont
  {Kokail}}, \bibinfo {author} {\bibfnamefont {C.}~\bibnamefont {Maier}},
  \bibinfo {author} {\bibfnamefont {R.}~\bibnamefont {van Bijnen}}, \bibinfo
  {author} {\bibfnamefont {T.}~\bibnamefont {Brydges}}, \bibinfo {author}
  {\bibfnamefont {M.~K.}\ \bibnamefont {Joshi}}, \bibinfo {author}
  {\bibfnamefont {P.}~\bibnamefont {Jurcevic}}, \bibinfo {author}
  {\bibfnamefont {C.~A.}\ \bibnamefont {Muschik}}, \bibinfo {author}
  {\bibfnamefont {P.}~\bibnamefont {Silvi}}, \bibinfo {author} {\bibfnamefont
  {R.}~\bibnamefont {Blatt}}, \bibinfo {author} {\bibfnamefont {C.~F.}\
  \bibnamefont {Roos}},  \emph {et~al.},\ }\bibfield  {title} {\enquote
  {\bibinfo {title} {Self-verifying variational quantum simulation of lattice
  models},}\ }\href {\doibase 10.1038/s41586-019-1177-4} {\bibfield  {journal}
  {\bibinfo  {journal} {Nature}\ }\textbf {\bibinfo {volume} {569}},\ \bibinfo
  {pages} {355} (\bibinfo {year} {2019}{\natexlab{a}})}\BibitemShut {NoStop}%
\bibitem [{\citenamefont {Schau\ss{}}\ \emph {et~al.}(2012)\citenamefont
  {Schau\ss{}}, \citenamefont {Cheneau}, \citenamefont {Endres}, \citenamefont
  {Fukuhara}, \citenamefont {Hild}, \citenamefont {Omran}, \citenamefont
  {Pohl}, \citenamefont {Gross}, \citenamefont {Kuhr},\ and\ \citenamefont
  {Bloch}}]{Schauss2012}%
  \BibitemOpen
  \bibfield  {author} {\bibinfo {author} {\bibfnamefont {P.}~\bibnamefont
  {Schau\ss{}}}, \bibinfo {author} {\bibfnamefont {M.}~\bibnamefont {Cheneau}},
  \bibinfo {author} {\bibfnamefont {M.}~\bibnamefont {Endres}}, \bibinfo
  {author} {\bibfnamefont {T.}~\bibnamefont {Fukuhara}}, \bibinfo {author}
  {\bibfnamefont {S.}~\bibnamefont {Hild}}, \bibinfo {author} {\bibfnamefont
  {A.}~\bibnamefont {Omran}}, \bibinfo {author} {\bibfnamefont
  {T.}~\bibnamefont {Pohl}}, \bibinfo {author} {\bibfnamefont {C.}~\bibnamefont
  {Gross}}, \bibinfo {author} {\bibfnamefont {S.}~\bibnamefont {Kuhr}}, \ and\
  \bibinfo {author} {\bibfnamefont {I.}~\bibnamefont {Bloch}},\ }\bibfield
  {title} {\enquote {\bibinfo {title} {Observation of spatially ordered
  structures in a two-dimensional {R}ydberg gas},}\ }\href {\doibase
  10.1038/nature11596} {\bibfield  {journal} {\bibinfo  {journal} {Nature}\
  }\textbf {\bibinfo {volume} {491}},\ \bibinfo {pages} {87--91} (\bibinfo
  {year} {2012})}\BibitemShut {NoStop}%
\bibitem [{\citenamefont {Labuhn}\ \emph {et~al.}(2016)\citenamefont {Labuhn},
  \citenamefont {Barredo}, \citenamefont {Ravets}, \citenamefont
  {De~L{\'e}s{\'e}leuc}, \citenamefont {Macr{\`\i}}, \citenamefont {Lahaye},\
  and\ \citenamefont {Browaeys}}]{labuhn2016}%
  \BibitemOpen
  \bibfield  {author} {\bibinfo {author} {\bibfnamefont {H.}~\bibnamefont
  {Labuhn}}, \bibinfo {author} {\bibfnamefont {D.}~\bibnamefont {Barredo}},
  \bibinfo {author} {\bibfnamefont {S.}~\bibnamefont {Ravets}}, \bibinfo
  {author} {\bibfnamefont {S.}~\bibnamefont {De~L{\'e}s{\'e}leuc}}, \bibinfo
  {author} {\bibfnamefont {T.}~\bibnamefont {Macr{\`\i}}}, \bibinfo {author}
  {\bibfnamefont {T.}~\bibnamefont {Lahaye}}, \ and\ \bibinfo {author}
  {\bibfnamefont {A.}~\bibnamefont {Browaeys}},\ }\bibfield  {title} {\enquote
  {\bibinfo {title} {Tunable two-dimensional arrays of single {R}ydberg atoms
  for realizing quantum {I}sing models},}\ }\href {\doibase
  10.1038/nature18274} {\bibfield  {journal} {\bibinfo  {journal} {Nature}\
  }\textbf {\bibinfo {volume} {534}},\ \bibinfo {pages} {667} (\bibinfo {year}
  {2016})}\BibitemShut {NoStop}%
\bibitem [{\citenamefont {Bernien}\ \emph {et~al.}(2017)\citenamefont
  {Bernien}, \citenamefont {Schwartz}, \citenamefont {Keesling}, \citenamefont
  {Levine}, \citenamefont {Omran}, \citenamefont {Pichler}, \citenamefont
  {Choi}, \citenamefont {Zibrov}, \citenamefont {Endres}, \citenamefont
  {Greiner} \emph {et~al.}}]{bernien2017}%
  \BibitemOpen
  \bibfield  {author} {\bibinfo {author} {\bibfnamefont {H.}~\bibnamefont
  {Bernien}}, \bibinfo {author} {\bibfnamefont {S.}~\bibnamefont {Schwartz}},
  \bibinfo {author} {\bibfnamefont {A.}~\bibnamefont {Keesling}}, \bibinfo
  {author} {\bibfnamefont {H.}~\bibnamefont {Levine}}, \bibinfo {author}
  {\bibfnamefont {A.}~\bibnamefont {Omran}}, \bibinfo {author} {\bibfnamefont
  {H.}~\bibnamefont {Pichler}}, \bibinfo {author} {\bibfnamefont
  {S.}~\bibnamefont {Choi}}, \bibinfo {author} {\bibfnamefont {A.~S.}\
  \bibnamefont {Zibrov}}, \bibinfo {author} {\bibfnamefont {M.}~\bibnamefont
  {Endres}}, \bibinfo {author} {\bibfnamefont {M.}~\bibnamefont {Greiner}},
  \emph {et~al.},\ }\bibfield  {title} {\enquote {\bibinfo {title} {Probing
  many-body dynamics on a 51-atom quantum simulator},}\ }\href {\doibase
  10.1038/nature24622} {\bibfield  {journal} {\bibinfo  {journal} {Nature}\
  }\textbf {\bibinfo {volume} {551}},\ \bibinfo {pages} {579} (\bibinfo {year}
  {2017})}\BibitemShut {NoStop}%
\bibitem [{\citenamefont {Guardado-Sanchez}\ \emph {et~al.}(2018)\citenamefont
  {Guardado-Sanchez}, \citenamefont {Brown}, \citenamefont {Mitra},
  \citenamefont {Devakul}, \citenamefont {Huse}, \citenamefont {Schau\ss{}},\
  and\ \citenamefont {Bakr}}]{Sanchez2018}%
  \BibitemOpen
  \bibfield  {author} {\bibinfo {author} {\bibfnamefont {E.}~\bibnamefont
  {Guardado-Sanchez}}, \bibinfo {author} {\bibfnamefont {P.~T.}\ \bibnamefont
  {Brown}}, \bibinfo {author} {\bibfnamefont {D.}~\bibnamefont {Mitra}},
  \bibinfo {author} {\bibfnamefont {T.}~\bibnamefont {Devakul}}, \bibinfo
  {author} {\bibfnamefont {D.~A.}\ \bibnamefont {Huse}}, \bibinfo {author}
  {\bibfnamefont {P.}~\bibnamefont {Schau\ss{}}}, \ and\ \bibinfo {author}
  {\bibfnamefont {W.~S.}\ \bibnamefont {Bakr}},\ }\bibfield  {title} {\enquote
  {\bibinfo {title} {Probing the quench dynamics of antiferromagnetic
  correlations in a {2D} quantum {I}sing spin system},}\ }\href {\doibase
  10.1103/PhysRevX.8.021069} {\bibfield  {journal} {\bibinfo  {journal} {Phys.
  Rev. X}\ }\textbf {\bibinfo {volume} {8}},\ \bibinfo {pages} {021069}
  (\bibinfo {year} {2018})}\BibitemShut {NoStop}%
\bibitem [{\citenamefont {Lienhard}\ \emph {et~al.}(2018)\citenamefont
  {Lienhard}, \citenamefont {{D}e L\'es\'eleuc}, \citenamefont {Barredo},
  \citenamefont {Lahaye}, \citenamefont {Browaeys}, \citenamefont {Schuler},
  \citenamefont {Henry},\ and\ \citenamefont {L\"auchli}}]{Lienhard2018}%
  \BibitemOpen
  \bibfield  {author} {\bibinfo {author} {\bibfnamefont {V.}~\bibnamefont
  {Lienhard}}, \bibinfo {author} {\bibfnamefont {S.}~\bibnamefont {{D}e
  L\'es\'eleuc}}, \bibinfo {author} {\bibfnamefont {D.}~\bibnamefont
  {Barredo}}, \bibinfo {author} {\bibfnamefont {T.}~\bibnamefont {Lahaye}},
  \bibinfo {author} {\bibfnamefont {A.}~\bibnamefont {Browaeys}}, \bibinfo
  {author} {\bibfnamefont {M.}~\bibnamefont {Schuler}}, \bibinfo {author}
  {\bibfnamefont {L.~P.}\ \bibnamefont {Henry}}, \ and\ \bibinfo {author}
  {\bibfnamefont {A.~M.}\ \bibnamefont {L\"auchli}},\ }\bibfield  {title}
  {\enquote {\bibinfo {title} {Observing the space- and time-dependent growth
  of correlations in dynamically tuned synthetic {I}sing models with
  antiferromagnetic interactions},}\ }\href {\doibase
  10.1103/PhysRevX.8.021070} {\bibfield  {journal} {\bibinfo  {journal} {Phys.
  Rev. X}\ }\textbf {\bibinfo {volume} {8}},\ \bibinfo {pages} {021070}
  (\bibinfo {year} {2018})}\BibitemShut {NoStop}%
\bibitem [{\citenamefont {M\"uller}\ \emph {et~al.}(2008)\citenamefont
  {M\"uller}, \citenamefont {Liang}, \citenamefont {Lesanovsky},\ and\
  \citenamefont {Zoller}}]{Muller_2008}%
  \BibitemOpen
  \bibfield  {author} {\bibinfo {author} {\bibfnamefont {M.}~\bibnamefont
  {M\"uller}}, \bibinfo {author} {\bibfnamefont {L.}~\bibnamefont {Liang}},
  \bibinfo {author} {\bibfnamefont {I.}~\bibnamefont {Lesanovsky}}, \ and\
  \bibinfo {author} {\bibfnamefont {P.}~\bibnamefont {Zoller}},\ }\bibfield
  {title} {\enquote {\bibinfo {title} {Trapped {R}ydberg ions: from spin chains
  to fast quantum gates},}\ }\href {\doibase 10.1088/1367-2630/10/9/093009}
  {\bibfield  {journal} {\bibinfo  {journal} {New J. Phys.}\ }\textbf {\bibinfo
  {volume} {10}},\ \bibinfo {pages} {093009} (\bibinfo {year}
  {2008})}\BibitemShut {NoStop}%
\bibitem [{\citenamefont {Zhang}\ \emph {et~al.}(2019)\citenamefont {Zhang},
  \citenamefont {Pokorny}, \citenamefont {Li}, \citenamefont {Higgins},
  \citenamefont {P\"oschl}, \citenamefont {Lesanovsky},\ and\ \citenamefont
  {Hennrich}}]{zhang2019}%
  \BibitemOpen
  \bibfield  {author} {\bibinfo {author} {\bibfnamefont {C.}~\bibnamefont
  {Zhang}}, \bibinfo {author} {\bibfnamefont {F.}~\bibnamefont {Pokorny}},
  \bibinfo {author} {\bibfnamefont {W.}~\bibnamefont {Li}}, \bibinfo {author}
  {\bibfnamefont {G.}~\bibnamefont {Higgins}}, \bibinfo {author} {\bibfnamefont
  {A.}~\bibnamefont {P\"oschl}}, \bibinfo {author} {\bibfnamefont
  {I.}~\bibnamefont {Lesanovsky}}, \ and\ \bibinfo {author} {\bibfnamefont
  {M.}~\bibnamefont {Hennrich}},\ }\bibfield  {title} {\enquote {\bibinfo
  {title} {Sub-microsecond entangling gate between trapped ions via {R}ydberg
  interaction},}\ }\href {https://arxiv.org/abs/1908.11284} {\bibfield
  {journal} {\bibinfo  {journal} {arXiv:1908.11284}\ } (\bibinfo {year}
  {2019})}\BibitemShut {NoStop}%
\bibitem [{\citenamefont {Whitlock}\ \emph {et~al.}(2017)\citenamefont
  {Whitlock}, \citenamefont {Glaetzle},\ and\ \citenamefont
  {Hannaford}}]{Whitlock_2017}%
  \BibitemOpen
  \bibfield  {author} {\bibinfo {author} {\bibfnamefont {S.}~\bibnamefont
  {Whitlock}}, \bibinfo {author} {\bibfnamefont {A.~W.}\ \bibnamefont
  {Glaetzle}}, \ and\ \bibinfo {author} {\bibfnamefont {P.}~\bibnamefont
  {Hannaford}},\ }\bibfield  {title} {\enquote {\bibinfo {title} {Simulating
  quantum spin models using {R}ydberg-excited atomic ensembles in magnetic
  microtrap arrays},}\ }\href {\doibase 10.1088/1361-6455/aa6149} {\bibfield
  {journal} {\bibinfo  {journal} {J. Phys. B}\ }\textbf {\bibinfo {volume}
  {50}},\ \bibinfo {pages} {074001} (\bibinfo {year} {2017})}\BibitemShut
  {NoStop}%
\bibitem [{\citenamefont {Letscher}\ \emph {et~al.}(2017)\citenamefont
  {Letscher}, \citenamefont {Petrosyan},\ and\ \citenamefont
  {Fleischhauer}}]{Letscher_2017}%
  \BibitemOpen
  \bibfield  {author} {\bibinfo {author} {\bibfnamefont {F.}~\bibnamefont
  {Letscher}}, \bibinfo {author} {\bibfnamefont {D.}~\bibnamefont {Petrosyan}},
  \ and\ \bibinfo {author} {\bibfnamefont {M.}~\bibnamefont {Fleischhauer}},\
  }\bibfield  {title} {\enquote {\bibinfo {title} {Many-body dynamics of holes
  in a driven, dissipative spin chain of {R}ydberg superatoms},}\ }\href
  {\doibase 10.1088/1367-2630/aa91c6} {\bibfield  {journal} {\bibinfo
  {journal} {New J. Phys.}\ }\textbf {\bibinfo {volume} {19}},\ \bibinfo
  {pages} {113014} (\bibinfo {year} {2017})}\BibitemShut {NoStop}%
\bibitem [{\citenamefont {Lesanovsky}(2011)}]{Lesanovsky2011}%
  \BibitemOpen
  \bibfield  {author} {\bibinfo {author} {\bibfnamefont {I.}~\bibnamefont
  {Lesanovsky}},\ }\bibfield  {title} {\enquote {\bibinfo {title} {Many-body
  spin interactions and the ground state of a dense {R}ydberg lattice gas},}\
  }\href {\doibase 10.1103/PhysRevLett.106.025301} {\bibfield  {journal}
  {\bibinfo  {journal} {Phys. Rev. Lett.}\ }\textbf {\bibinfo {volume} {106}},\
  \bibinfo {pages} {025301} (\bibinfo {year} {2011})}\BibitemShut {NoStop}%
\bibitem [{\citenamefont {Sanz}\ \emph {et~al.}(2016)\citenamefont {Sanz},
  \citenamefont {Solano},\ and\ \citenamefont {Egusquiza}}]{Sanz2016}%
  \BibitemOpen
  \bibfield  {author} {\bibinfo {author} {\bibfnamefont {M.}~\bibnamefont
  {Sanz}}, \bibinfo {author} {\bibfnamefont {E.}~\bibnamefont {Solano}}, \ and\
  \bibinfo {author} {\bibfnamefont {{\'I}.~L.}\ \bibnamefont {Egusquiza}},\
  }\bibfield  {title} {\enquote {\bibinfo {title} {Beyond adiabatic
  elimination: Effective {H}amiltonians and singular perturbation},}\ }in\
  \href {\doibase 10.1007/978-4-431-55342-7} {\emph {\bibinfo {booktitle}
  {Applications + Practical Conceptualization + Mathematics = fruitful
  Innovation}}}\ (\bibinfo  {publisher} {Springer Japan},\ \bibinfo {address}
  {Tokyo},\ \bibinfo {year} {2016})\ pp.\ \bibinfo {pages}
  {127--142}\BibitemShut {NoStop}%
\bibitem [{\citenamefont {James}\ and\ \citenamefont
  {Jerke}(2007)}]{james2007effective}%
  \BibitemOpen
  \bibfield  {author} {\bibinfo {author} {\bibfnamefont {D.~F.}\ \bibnamefont
  {James}}\ and\ \bibinfo {author} {\bibfnamefont {J.}~\bibnamefont {Jerke}},\
  }\bibfield  {title} {\enquote {\bibinfo {title} {Effective {H}amiltonian
  theory and its applications in quantum information},}\ }\href {\doibase
  10.1139/p07-060} {\bibfield  {journal} {\bibinfo  {journal} {Can. J. Phys}\
  }\textbf {\bibinfo {volume} {85}},\ \bibinfo {pages} {625--632} (\bibinfo
  {year} {2007})}\BibitemShut {NoStop}%
\bibitem [{\citenamefont {Reiter}\ and\ \citenamefont
  {S\o{}rensen}(2012)}]{Reiter2012}%
  \BibitemOpen
  \bibfield  {author} {\bibinfo {author} {\bibfnamefont {F.}~\bibnamefont
  {Reiter}}\ and\ \bibinfo {author} {\bibfnamefont {A.~S.}\ \bibnamefont
  {S\o{}rensen}},\ }\bibfield  {title} {\enquote {\bibinfo {title} {Effective
  operator formalism for open quantum systems},}\ }\href {\doibase
  10.1103/PhysRevA.85.032111} {\bibfield  {journal} {\bibinfo  {journal} {Phys.
  Rev. A}\ }\textbf {\bibinfo {volume} {85}},\ \bibinfo {pages} {032111}
  (\bibinfo {year} {2012})}\BibitemShut {NoStop}%
\bibitem [{\citenamefont {Olmos}\ \emph {et~al.}(2012)\citenamefont {Olmos},
  \citenamefont {Lesanovsky},\ and\ \citenamefont {Garrahan}}]{Olmos2012facil}%
  \BibitemOpen
  \bibfield  {author} {\bibinfo {author} {\bibfnamefont {B.}~\bibnamefont
  {Olmos}}, \bibinfo {author} {\bibfnamefont {I.}~\bibnamefont {Lesanovsky}}, \
  and\ \bibinfo {author} {\bibfnamefont {J.~P.}\ \bibnamefont {Garrahan}},\
  }\bibfield  {title} {\enquote {\bibinfo {title} {Facilitated spin models of
  dissipative quantum glasses},}\ }\href {\doibase
  10.1103/PhysRevLett.109.020403} {\bibfield  {journal} {\bibinfo  {journal}
  {Phys. Rev. Lett.}\ }\textbf {\bibinfo {volume} {109}},\ \bibinfo {pages}
  {020403} (\bibinfo {year} {2012})}\BibitemShut {NoStop}%
\bibitem [{\citenamefont {Turner}\ \emph {et~al.}(2018)\citenamefont {Turner},
  \citenamefont {Michailidis}, \citenamefont {Abanin}, \citenamefont {Serbyn},\
  and\ \citenamefont {Papi\ifmmode~\acute{c}\else \'{c}\fi{}}}]{Turner2018}%
  \BibitemOpen
  \bibfield  {author} {\bibinfo {author} {\bibfnamefont {C.~J.}\ \bibnamefont
  {Turner}}, \bibinfo {author} {\bibfnamefont {A.~A.}\ \bibnamefont
  {Michailidis}}, \bibinfo {author} {\bibfnamefont {D.~A.}\ \bibnamefont
  {Abanin}}, \bibinfo {author} {\bibfnamefont {M.}~\bibnamefont {Serbyn}}, \
  and\ \bibinfo {author} {\bibfnamefont {Z.}~\bibnamefont
  {Papi\ifmmode~\acute{c}\else \'{c}\fi{}}},\ }\bibfield  {title} {\enquote
  {\bibinfo {title} {Quantum scarred eigenstates in a {R}ydberg atom chain:
  Entanglement, breakdown of thermalization, and stability to perturbations},}\
  }\href {\doibase 10.1103/PhysRevB.98.155134} {\bibfield  {journal} {\bibinfo
  {journal} {Phys. Rev. B}\ }\textbf {\bibinfo {volume} {98}},\ \bibinfo
  {pages} {155134} (\bibinfo {year} {2018})}\BibitemShut {NoStop}%
\bibitem [{\citenamefont {Lesanovsky}\ \emph {et~al.}(2019)\citenamefont
  {Lesanovsky}, \citenamefont {Macieszczak},\ and\ \citenamefont
  {Garrahan}}]{Lesanovsky2019}%
  \BibitemOpen
  \bibfield  {author} {\bibinfo {author} {\bibfnamefont {I.}~\bibnamefont
  {Lesanovsky}}, \bibinfo {author} {\bibfnamefont {K.}~\bibnamefont
  {Macieszczak}}, \ and\ \bibinfo {author} {\bibfnamefont {J.~P.}\ \bibnamefont
  {Garrahan}},\ }\bibfield  {title} {\enquote {\bibinfo {title}
  {Non-equilibrium absorbing state phase transitions in discrete-time quantum
  cellular automaton dynamics on spin lattices},}\ }\href {\doibase
  10.1088/2058-9565/aaf831} {\bibfield  {journal} {\bibinfo  {journal} {Quantum
  Sci. Technol.}\ }\textbf {\bibinfo {volume} {4}},\ \bibinfo {pages} {02LT02}
  (\bibinfo {year} {2019})}\BibitemShut {NoStop}%
\bibitem [{\citenamefont {Ostmann}\ \emph {et~al.}(2019)\citenamefont
  {Ostmann}, \citenamefont {Marcuzzi}, \citenamefont {Garrahan},\ and\
  \citenamefont {Lesanovsky}}]{ostmann2019}%
  \BibitemOpen
  \bibfield  {author} {\bibinfo {author} {\bibfnamefont {M.}~\bibnamefont
  {Ostmann}}, \bibinfo {author} {\bibfnamefont {M.}~\bibnamefont {Marcuzzi}},
  \bibinfo {author} {\bibfnamefont {J.~P.}\ \bibnamefont {Garrahan}}, \ and\
  \bibinfo {author} {\bibfnamefont {I.}~\bibnamefont {Lesanovsky}},\ }\bibfield
   {title} {\enquote {\bibinfo {title} {Localization in spin chains with
  facilitation constraints and disordered interactions},}\ }\href {\doibase
  10.1103/PhysRevA.99.060101} {\bibfield  {journal} {\bibinfo  {journal} {Phys.
  Rev. A}\ }\textbf {\bibinfo {volume} {99}},\ \bibinfo {pages} {060101}
  (\bibinfo {year} {2019})}\BibitemShut {NoStop}%
\bibitem [{\citenamefont {Johansson}\ \emph {et~al.}(2013)\citenamefont
  {Johansson}, \citenamefont {Nation},\ and\ \citenamefont {Nori}}]{qutip2013}%
  \BibitemOpen
  \bibfield  {author} {\bibinfo {author} {\bibfnamefont {J.~R.}\ \bibnamefont
  {Johansson}}, \bibinfo {author} {\bibfnamefont {P.~D.}\ \bibnamefont
  {Nation}}, \ and\ \bibinfo {author} {\bibfnamefont {F.}~\bibnamefont
  {Nori}},\ }\bibfield  {title} {\enquote {\bibinfo {title} {Qutip 2: A
  {Python} framework for the dynamics of open quantum systems},}\ }\href
  {\doibase https://doi.org/10.1016/j.cpc.2012.11.019} {\bibfield  {journal}
  {\bibinfo  {journal} {Comput. Phys. Commun.}\ }\textbf {\bibinfo {volume}
  {184}},\ \bibinfo {pages} {1234 -- 1240} (\bibinfo {year}
  {2013})}\BibitemShut {NoStop}%
\bibitem [{\citenamefont {Hillberry}(2016)}]{hillberry2016}%
  \BibitemOpen
  \bibfield  {author} {\bibinfo {author} {\bibfnamefont {L.~E.}\ \bibnamefont
  {Hillberry}},\ }\emph {\bibinfo {title} {Entanglement and complexity in
  quantum elementary cellular automata}},\ \href {\doibase 11124/170336} {Ph.D.
  thesis},\ \bibinfo  {school} {Colorado School of Mines. Arthur Lakes Library}
  (\bibinfo {year} {2016})\BibitemShut {NoStop}%
\bibitem [{\citenamefont {Peruzzo}\ \emph {et~al.}(2014)\citenamefont
  {Peruzzo}, \citenamefont {McClean}, \citenamefont {Shadbolt}, \citenamefont
  {Yung}, \citenamefont {Zhou}, \citenamefont {Love}, \citenamefont
  {Aspuru-Guzik},\ and\ \citenamefont {O'brien}}]{peruzzo2014variational}%
  \BibitemOpen
  \bibfield  {author} {\bibinfo {author} {\bibfnamefont {A.}~\bibnamefont
  {Peruzzo}}, \bibinfo {author} {\bibfnamefont {J.}~\bibnamefont {McClean}},
  \bibinfo {author} {\bibfnamefont {P.}~\bibnamefont {Shadbolt}}, \bibinfo
  {author} {\bibfnamefont {M.~H.}\ \bibnamefont {Yung}}, \bibinfo {author}
  {\bibfnamefont {X.~Q.}\ \bibnamefont {Zhou}}, \bibinfo {author}
  {\bibfnamefont {P.~J.}\ \bibnamefont {Love}}, \bibinfo {author}
  {\bibfnamefont {A.}~\bibnamefont {Aspuru-Guzik}}, \ and\ \bibinfo {author}
  {\bibfnamefont {J.~L.}\ \bibnamefont {O'brien}},\ }\bibfield  {title}
  {\enquote {\bibinfo {title} {A variational eigenvalue solver on a photonic
  quantum processor},}\ }\href {\doibase 10.1038/ncomms5213} {\bibfield
  {journal} {\bibinfo  {journal} {Nat. Commun.}\ }\textbf {\bibinfo {volume}
  {5}},\ \bibinfo {pages} {4213} (\bibinfo {year} {2014})}\BibitemShut
  {NoStop}%
\bibitem [{\citenamefont {O'Malley}\ \emph {et~al.}(2016)\citenamefont
  {O'Malley}, \citenamefont {Babbush}, \citenamefont {Kivlichan}, \citenamefont
  {Romero}, \citenamefont {McClean}, \citenamefont {Barends}, \citenamefont
  {Kelly}, \citenamefont {Roushan}, \citenamefont {Tranter}, \citenamefont
  {Ding} \emph {et~al.}}]{omalley2016scalable}%
  \BibitemOpen
  \bibfield  {author} {\bibinfo {author} {\bibfnamefont {P.~J.~J.}\
  \bibnamefont {O'Malley}}, \bibinfo {author} {\bibfnamefont {R.}~\bibnamefont
  {Babbush}}, \bibinfo {author} {\bibfnamefont {I.~D.}\ \bibnamefont
  {Kivlichan}}, \bibinfo {author} {\bibfnamefont {J.}~\bibnamefont {Romero}},
  \bibinfo {author} {\bibfnamefont {J.~R.}\ \bibnamefont {McClean}}, \bibinfo
  {author} {\bibfnamefont {R.}~\bibnamefont {Barends}}, \bibinfo {author}
  {\bibfnamefont {J.}~\bibnamefont {Kelly}}, \bibinfo {author} {\bibfnamefont
  {P.}~\bibnamefont {Roushan}}, \bibinfo {author} {\bibfnamefont
  {A.}~\bibnamefont {Tranter}}, \bibinfo {author} {\bibfnamefont
  {N.}~\bibnamefont {Ding}},  \emph {et~al.},\ }\bibfield  {title} {\enquote
  {\bibinfo {title} {Scalable quantum simulation of molecular energies},}\
  }\href {\doibase 10.1103/PhysRevX.6.031007} {\bibfield  {journal} {\bibinfo
  {journal} {Phys. Rev. X}\ }\textbf {\bibinfo {volume} {6}},\ \bibinfo {pages}
  {031007} (\bibinfo {year} {2016})}\BibitemShut {NoStop}%
\bibitem [{\citenamefont {Kandala}\ \emph {et~al.}(2017)\citenamefont
  {Kandala}, \citenamefont {Mezzacapo}, \citenamefont {Temme}, \citenamefont
  {Takita}, \citenamefont {Brink}, \citenamefont {Chow},\ and\ \citenamefont
  {Gambetta}}]{kandala2017hardware}%
  \BibitemOpen
  \bibfield  {author} {\bibinfo {author} {\bibfnamefont {A.}~\bibnamefont
  {Kandala}}, \bibinfo {author} {\bibfnamefont {A.}~\bibnamefont {Mezzacapo}},
  \bibinfo {author} {\bibfnamefont {K.}~\bibnamefont {Temme}}, \bibinfo
  {author} {\bibfnamefont {M.}~\bibnamefont {Takita}}, \bibinfo {author}
  {\bibfnamefont {M.}~\bibnamefont {Brink}}, \bibinfo {author} {\bibfnamefont
  {J.~M.}\ \bibnamefont {Chow}}, \ and\ \bibinfo {author} {\bibfnamefont
  {J.~M.}\ \bibnamefont {Gambetta}},\ }\bibfield  {title} {\enquote {\bibinfo
  {title} {Hardware-efficient variational quantum eigensolver for small
  molecules and quantum magnets},}\ }\href {\doibase 10.1038/nature23879}
  {\bibfield  {journal} {\bibinfo  {journal} {Nature}\ }\textbf {\bibinfo
  {volume} {549}},\ \bibinfo {pages} {242} (\bibinfo {year}
  {2017})}\BibitemShut {NoStop}%
\bibitem [{\citenamefont {Kokail}\ \emph
  {et~al.}(2019{\natexlab{b}})\citenamefont {Kokail}, \citenamefont {Maier},
  \citenamefont {van Bijnen}, \citenamefont {Brydges}, \citenamefont {Joshi},
  \citenamefont {Jurcevic}, \citenamefont {Muschik}, \citenamefont {Silvi},
  \citenamefont {Blatt}, \citenamefont {Roos} \emph {et~al.}}]{kokail2019self}%
  \BibitemOpen
  \bibfield  {author} {\bibinfo {author} {\bibfnamefont {C.}~\bibnamefont
  {Kokail}}, \bibinfo {author} {\bibfnamefont {C.}~\bibnamefont {Maier}},
  \bibinfo {author} {\bibfnamefont {R.}~\bibnamefont {van Bijnen}}, \bibinfo
  {author} {\bibfnamefont {T.}~\bibnamefont {Brydges}}, \bibinfo {author}
  {\bibfnamefont {M.~K.}\ \bibnamefont {Joshi}}, \bibinfo {author}
  {\bibfnamefont {P.}~\bibnamefont {Jurcevic}}, \bibinfo {author}
  {\bibfnamefont {C.~A.}\ \bibnamefont {Muschik}}, \bibinfo {author}
  {\bibfnamefont {P.}~\bibnamefont {Silvi}}, \bibinfo {author} {\bibfnamefont
  {R.}~\bibnamefont {Blatt}}, \bibinfo {author} {\bibfnamefont {C.~F.}\
  \bibnamefont {Roos}},  \emph {et~al.},\ }\bibfield  {title} {\enquote
  {\bibinfo {title} {Self-verifying variational quantum simulation of lattice
  models},}\ }\href {\doibase 10.1038/s41586-019-1177-4} {\bibfield  {journal}
  {\bibinfo  {journal} {Nature}\ }\textbf {\bibinfo {volume} {569}},\ \bibinfo
  {pages} {355} (\bibinfo {year} {2019}{\natexlab{b}})}\BibitemShut {NoStop}%
\bibitem [{\citenamefont {Kennedy}(2010)}]{kennedy2010particle}%
  \BibitemOpen
  \bibfield  {author} {\bibinfo {author} {\bibfnamefont {J.}~\bibnamefont
  {Kennedy}},\ }\bibfield  {title} {\enquote {\bibinfo {title} {Particle swarm
  optimization},}\ }\href {\doibase 10.1007/978-0-387-30164-8_630} {\bibfield
  {journal} {\bibinfo  {journal} {Encyclopedia of machine learning}\ ,\
  \bibinfo {pages} {760--766}} (\bibinfo {year} {2010})}\BibitemShut {NoStop}%
\bibitem [{\citenamefont {Omran}\ \emph {et~al.}(2019)\citenamefont {Omran},
  \citenamefont {Levine}, \citenamefont {Keesling}, \citenamefont {Semeghini},
  \citenamefont {Wang}, \citenamefont {Ebadi}, \citenamefont {Bernien},
  \citenamefont {Zibrov}, \citenamefont {Pichler}, \citenamefont {Choi},
  \citenamefont {Cui}, \citenamefont {Rossignolo}, \citenamefont {Rembold},
  \citenamefont {Montangero}, \citenamefont {Calarco}, \citenamefont {Endres},
  \citenamefont {Greiner}, \citenamefont {Vuleti{\'c}},\ and\ \citenamefont
  {Lukin}}]{Omran570}%
  \BibitemOpen
  \bibfield  {author} {\bibinfo {author} {\bibfnamefont {A.}~\bibnamefont
  {Omran}}, \bibinfo {author} {\bibfnamefont {H.}~\bibnamefont {Levine}},
  \bibinfo {author} {\bibfnamefont {A.}~\bibnamefont {Keesling}}, \bibinfo
  {author} {\bibfnamefont {G.}~\bibnamefont {Semeghini}}, \bibinfo {author}
  {\bibfnamefont {T.~T.}\ \bibnamefont {Wang}}, \bibinfo {author}
  {\bibfnamefont {S.}~\bibnamefont {Ebadi}}, \bibinfo {author} {\bibfnamefont
  {H.}~\bibnamefont {Bernien}}, \bibinfo {author} {\bibfnamefont {A.~S.}\
  \bibnamefont {Zibrov}}, \bibinfo {author} {\bibfnamefont {H.}~\bibnamefont
  {Pichler}}, \bibinfo {author} {\bibfnamefont {S.}~\bibnamefont {Choi}},
  \bibinfo {author} {\bibfnamefont {J.}~\bibnamefont {Cui}}, \bibinfo {author}
  {\bibfnamefont {M.}~\bibnamefont {Rossignolo}}, \bibinfo {author}
  {\bibfnamefont {P.}~\bibnamefont {Rembold}}, \bibinfo {author} {\bibfnamefont
  {S.}~\bibnamefont {Montangero}}, \bibinfo {author} {\bibfnamefont
  {T.}~\bibnamefont {Calarco}}, \bibinfo {author} {\bibfnamefont
  {M.}~\bibnamefont {Endres}}, \bibinfo {author} {\bibfnamefont
  {M.}~\bibnamefont {Greiner}}, \bibinfo {author} {\bibfnamefont
  {V.}~\bibnamefont {Vuleti{\'c}}}, \ and\ \bibinfo {author} {\bibfnamefont
  {M.~D.}\ \bibnamefont {Lukin}},\ }\bibfield  {title} {\enquote {\bibinfo
  {title} {Generation and manipulation of {S}chr{\"o}dinger cat states in
  {R}ydberg atom arrays},}\ }\href {\doibase 10.1126/science.aax9743}
  {\bibfield  {journal} {\bibinfo  {journal} {Science}\ }\textbf {\bibinfo
  {volume} {365}},\ \bibinfo {pages} {570--574} (\bibinfo {year}
  {2019})}\BibitemShut {NoStop}%
\bibitem [{\citenamefont {D{\"u}r}\ \emph {et~al.}(2014)\citenamefont
  {D{\"u}r}, \citenamefont {Skotiniotis}, \citenamefont {Froewis},\ and\
  \citenamefont {Kraus}}]{dur2014improved}%
  \BibitemOpen
  \bibfield  {author} {\bibinfo {author} {\bibfnamefont {W.}~\bibnamefont
  {D{\"u}r}}, \bibinfo {author} {\bibfnamefont {M.}~\bibnamefont
  {Skotiniotis}}, \bibinfo {author} {\bibfnamefont {F.}~\bibnamefont
  {Froewis}}, \ and\ \bibinfo {author} {\bibfnamefont {B.}~\bibnamefont
  {Kraus}},\ }\bibfield  {title} {\enquote {\bibinfo {title} {Improved quantum
  metrology using quantum error correction},}\ }\href {\doibase
  10.1103/PhysRevLett.112.080801} {\bibfield  {journal} {\bibinfo  {journal}
  {Phys. Rev. Lett.}\ }\textbf {\bibinfo {volume} {112}},\ \bibinfo {pages}
  {080801} (\bibinfo {year} {2014})}\BibitemShut {NoStop}%
\bibitem [{\citenamefont {Nickerson}\ \emph {et~al.}(2013)\citenamefont
  {Nickerson}, \citenamefont {Li},\ and\ \citenamefont
  {Benjamin}}]{nickerson2013topological}%
  \BibitemOpen
  \bibfield  {author} {\bibinfo {author} {\bibfnamefont {N.~H.}\ \bibnamefont
  {Nickerson}}, \bibinfo {author} {\bibfnamefont {Y.}~\bibnamefont {Li}}, \
  and\ \bibinfo {author} {\bibfnamefont {S.~C.}\ \bibnamefont {Benjamin}},\
  }\bibfield  {title} {\enquote {\bibinfo {title} {Topological quantum
  computing with a very noisy network and local error rates approaching one
  percent},}\ }\href {\doibase 10.1038/ncomms2773} {\bibfield  {journal}
  {\bibinfo  {journal} {Nat. Commun.}\ }\textbf {\bibinfo {volume} {4}},\
  \bibinfo {pages} {1756} (\bibinfo {year} {2013})}\BibitemShut {NoStop}%
\bibitem [{\citenamefont {Barredo}\ \emph {et~al.}(2019)\citenamefont
  {Barredo}, \citenamefont {Lienhard}, \citenamefont {Scholl}, \citenamefont
  {de~L{\'e}s{\'e}leuc}, \citenamefont {Boulier}, \citenamefont {Browaeys},\
  and\ \citenamefont {Lahaye}}]{barredo2019three}%
  \BibitemOpen
  \bibfield  {author} {\bibinfo {author} {\bibfnamefont {D.}~\bibnamefont
  {Barredo}}, \bibinfo {author} {\bibfnamefont {V.}~\bibnamefont {Lienhard}},
  \bibinfo {author} {\bibfnamefont {P.}~\bibnamefont {Scholl}}, \bibinfo
  {author} {\bibfnamefont {S.}~\bibnamefont {de~L{\'e}s{\'e}leuc}}, \bibinfo
  {author} {\bibfnamefont {T.}~\bibnamefont {Boulier}}, \bibinfo {author}
  {\bibfnamefont {A.}~\bibnamefont {Browaeys}}, \ and\ \bibinfo {author}
  {\bibfnamefont {T.}~\bibnamefont {Lahaye}},\ }\bibfield  {title} {\enquote
  {\bibinfo {title} {Three-dimensional trapping of individual {R}ydberg atoms
  in ponderomotive bottle beam traps},}\ }\href
  {https://arxiv.org/abs/1908.00853} {\bibfield  {journal} {\bibinfo  {journal}
  {arXiv:1908.00853}\ } (\bibinfo {year} {2019})}\BibitemShut {NoStop}%
\bibitem [{\citenamefont {Graham}\ \emph {et~al.}(2019)\citenamefont {Graham},
  \citenamefont {Kwon}, \citenamefont {Grinkemeyer}, \citenamefont {Marra},
  \citenamefont {Jiang}, \citenamefont {Lichtman}, \citenamefont {Sun},
  \citenamefont {Ebert},\ and\ \citenamefont {Saffman}}]{graham2019rydberg}%
  \BibitemOpen
  \bibfield  {author} {\bibinfo {author} {\bibfnamefont {T.~M.}\ \bibnamefont
  {Graham}}, \bibinfo {author} {\bibfnamefont {M.}~\bibnamefont {Kwon}},
  \bibinfo {author} {\bibfnamefont {B.}~\bibnamefont {Grinkemeyer}}, \bibinfo
  {author} {\bibfnamefont {Z.}~\bibnamefont {Marra}}, \bibinfo {author}
  {\bibfnamefont {X.}~\bibnamefont {Jiang}}, \bibinfo {author} {\bibfnamefont
  {M.~T.}\ \bibnamefont {Lichtman}}, \bibinfo {author} {\bibfnamefont
  {Y.}~\bibnamefont {Sun}}, \bibinfo {author} {\bibfnamefont {M.}~\bibnamefont
  {Ebert}}, \ and\ \bibinfo {author} {\bibfnamefont {M.}~\bibnamefont
  {Saffman}},\ }\bibfield  {title} {\enquote {\bibinfo {title}
  {{R}ydberg-mediated entanglement in a two-dimensional neutral atom qubit
  array},}\ }\href {\doibase 10.1103/PhysRevLett.123.230501} {\bibfield
  {journal} {\bibinfo  {journal} {Phys. Rev. Lett.}\ }\textbf {\bibinfo
  {volume} {123}},\ \bibinfo {pages} {230501} (\bibinfo {year}
  {2019})}\BibitemShut {NoStop}%
\bibitem [{\citenamefont {Kaufman}\ \emph {et~al.}(2012)\citenamefont
  {Kaufman}, \citenamefont {Lester},\ and\ \citenamefont
  {Regal}}]{Kaufman2012}%
  \BibitemOpen
  \bibfield  {author} {\bibinfo {author} {\bibfnamefont {A.~M.}\ \bibnamefont
  {Kaufman}}, \bibinfo {author} {\bibfnamefont {B.~J.}\ \bibnamefont {Lester}},
  \ and\ \bibinfo {author} {\bibfnamefont {C.~A.}\ \bibnamefont {Regal}},\
  }\bibfield  {title} {\enquote {\bibinfo {title} {Cooling a single atom in an
  optical tweezer to its quantum ground state},}\ }\href {\doibase
  10.1103/PhysRevX.2.041014} {\bibfield  {journal} {\bibinfo  {journal} {Phys.
  Rev. X}\ }\textbf {\bibinfo {volume} {2}},\ \bibinfo {pages} {041014}
  (\bibinfo {year} {2012})}\BibitemShut {NoStop}%
\bibitem [{\citenamefont {Fitzsimons}\ and\ \citenamefont
  {Twamley}(2009)}]{fitzsimons2009quantum}%
  \BibitemOpen
  \bibfield  {author} {\bibinfo {author} {\bibfnamefont {J.}~\bibnamefont
  {Fitzsimons}}\ and\ \bibinfo {author} {\bibfnamefont {J.}~\bibnamefont
  {Twamley}},\ }\bibfield  {title} {\enquote {\bibinfo {title} {Quantum fault
  tolerance in systems with restricted control},}\ }\href {\doibase
  10.1016/j.entcs.2009.12.012} {\bibfield  {journal} {\bibinfo  {journal}
  {Electron. Notes Theor. Comput. Sci.}\ }\textbf {\bibinfo {volume} {258}},\
  \bibinfo {pages} {35} (\bibinfo {year} {2009})}\BibitemShut {NoStop}%
\bibitem [{\citenamefont {Paz-Silva}\ \emph {et~al.}(2011)\citenamefont
  {Paz-Silva}, \citenamefont {Brennen},\ and\ \citenamefont
  {Twamley}}]{paz2011bulk}%
  \BibitemOpen
  \bibfield  {author} {\bibinfo {author} {\bibfnamefont {G.~A.}\ \bibnamefont
  {Paz-Silva}}, \bibinfo {author} {\bibfnamefont {G.~K.}\ \bibnamefont
  {Brennen}}, \ and\ \bibinfo {author} {\bibfnamefont {J.}~\bibnamefont
  {Twamley}},\ }\bibfield  {title} {\enquote {\bibinfo {title} {Bulk
  fault-tolerant quantum information processing with boundary
  addressability},}\ }\href {\doibase 10.1088/1367-2630/13/1/013011} {\bibfield
   {journal} {\bibinfo  {journal} {New J. Phys.}\ }\textbf {\bibinfo {volume}
  {13}},\ \bibinfo {pages} {013011} (\bibinfo {year} {2011})}\BibitemShut
  {NoStop}%
\bibitem [{\citenamefont {Nguyen}\ \emph {et~al.}(2018)\citenamefont {Nguyen},
  \citenamefont {Raimond}, \citenamefont {Sayrin}, \citenamefont {Cortinas},
  \citenamefont {Cantat-Moltrecht}, \citenamefont {Assemat}, \citenamefont
  {Dotsenko}, \citenamefont {Gleyzes}, \citenamefont {Haroche}, \citenamefont
  {Roux}, \citenamefont {Jolicoeur},\ and\ \citenamefont
  {Brune}}]{nguyen2018towards}%
  \BibitemOpen
  \bibfield  {author} {\bibinfo {author} {\bibfnamefont {T.~L.}\ \bibnamefont
  {Nguyen}}, \bibinfo {author} {\bibfnamefont {J.~M.}\ \bibnamefont {Raimond}},
  \bibinfo {author} {\bibfnamefont {C.}~\bibnamefont {Sayrin}}, \bibinfo
  {author} {\bibfnamefont {R.}~\bibnamefont {Cortinas}}, \bibinfo {author}
  {\bibfnamefont {T.}~\bibnamefont {Cantat-Moltrecht}}, \bibinfo {author}
  {\bibfnamefont {F.}~\bibnamefont {Assemat}}, \bibinfo {author} {\bibfnamefont
  {I.}~\bibnamefont {Dotsenko}}, \bibinfo {author} {\bibfnamefont
  {S.}~\bibnamefont {Gleyzes}}, \bibinfo {author} {\bibfnamefont
  {S.}~\bibnamefont {Haroche}}, \bibinfo {author} {\bibfnamefont
  {G.}~\bibnamefont {Roux}}, \bibinfo {author} {\bibfnamefont {Th.}\
  \bibnamefont {Jolicoeur}}, \ and\ \bibinfo {author} {\bibfnamefont
  {M.}~\bibnamefont {Brune}},\ }\bibfield  {title} {\enquote {\bibinfo {title}
  {Towards quantum simulation with circular {R}ydberg atoms},}\ }\href
  {\doibase 10.1103/PhysRevX.8.011032} {\bibfield  {journal} {\bibinfo
  {journal} {Phys. Rev. X}\ }\textbf {\bibinfo {volume} {8}},\ \bibinfo {pages}
  {011032} (\bibinfo {year} {2018})}\BibitemShut {NoStop}%
\end{thebibliography}%

\clearpage


\section*{Supplementary material}

# Supplemental Material
# Unitary and non-unitary quantum cellular automata with Rydberg arrays

T. M. Wintermantel,[1,2] Y. Wang,[2] G. Lochead,[2] S. Shevate,[2] G. K. Brennen,[3] and S. Whitlock[2]

[1]*Physikalisches Institut, Universität Heidelberg, Im Neuenheimer Feld 226, 69120 Heidelberg, Germany*
[2]*ISIS (UMR 7006) and IPCMS (UMR 7504), University of Strasbourg and CNRS, 67000 Strasbourg, France*
[3]*Center for Engineered Quantum Systems, Dept. of Physics & Astronomy, Macquarie University, 2109 NSW, Australia*
(Dated: January 30, 2020)

## DERIVATION OF THE EFFECTIVE TWO-LEVEL HAMILTONIAN

To derive the effective time-independent master equation [Eqs. (2) and (3) in the manuscript] from the full time-dependent Hamiltonian [Eq. (1)] we assume that the nearest-neighbor interaction $V$ and the intermediate state decay rate $\Gamma$ are the dominant scales. This makes it possible to neglect the non-resonant couplings of the driving field and to adiabatically eliminate the $|e\rangle$ state. In the following we use units where $\hbar = 1$.

We consider a multifrequency coupling field of the form $\mathcal{E}_j = \mathcal{E}_j^\theta + \mathcal{E}_j^\phi + c.c.$, where

$$\mathcal{E}_j^\theta = \frac{1}{2} \sum_k \theta_j^k e^{iE_r t + ikVt} \qquad (1)$$

$$\mathcal{E}_j^\phi = \frac{1}{2} \sum_k \phi_j^k e^{i(E_r - E_e)t + ikVt}, \qquad (2)$$

acting on the time-dependent Hamiltonian

$$\hat{H}(t) = \sum_j \underbrace{(\hat{\sigma}_j^{gr} \mathcal{E}_j^\theta(t) + \hat{\sigma}_j^{er} \mathcal{E}_j^\phi(t) + h.c.) + \hat{V}_{\rm int}}_{\hat{H}_{\rm int}} + \hat{H}_{\rm atom}, \qquad (3)$$

where $\hat{\sigma}^{ab} = |a\rangle\langle b|$, $\hat{H}_{\rm atom} = E_r \hat{\sigma}_j^{rr} + E_e \hat{\sigma}_j^{ee}$ and we restrict to nearest neighbor interactions between sites $j$ only $\hat{V}_{\rm int} = V \hat{\sigma}_j^{rr} \hat{\sigma}_{j+1}^{rr}$.

The first step is to transform away the time dependence due to the carrier frequencies $E_r, (E_r - E_e)$. We transform to a rotating frame via the unitary $\hat{H}' = \hat{U}^\dagger \hat{H}_{\rm int} \hat{U}$ where $\hat{U} = \exp(-i\hat{H}_{\rm atom} t)$ and average out the rapidly varying phases that depend on $E_r, E_e$.

The resulting (still time-dependent) Hamiltonian in the rotating wave approximation is

$$\hat{H}(t) = \sum_j \sum_k \left( \frac{\theta_j^k}{2} \hat{\sigma}_j^{gr} e^{ikVt} + \frac{(\theta_j^k)^*}{2} \hat{\sigma}_j^{rg} e^{-ikVt} \right. \qquad (4)$$
$$\left. + \frac{\phi_j^k}{2} \hat{\sigma}_j^{er} e^{ikVt} + \frac{(\phi_j^k)^*}{2} \hat{\sigma}_j^{re} e^{-ikVt} \right) + V \hat{\sigma}_j^{rr} \hat{\sigma}_{j+1}^{rr}.$$

Next we transform to an interaction picture with respect to the nearest neighbor interaction using the unitary transformation $U = \exp\left(-iVt \sum_j \hat{\sigma}_j^{rr} \hat{\sigma}_{j+1}^{rr}\right)$ where $U^\dagger \hat{\sigma}^{\alpha r} U = [P_{j-1}^0 + P_{j-1}^1 e^{-iVt}] \hat{\sigma}_j^{\alpha r} [P_{j+1}^0 + P_{j+1}^1 e^{-iVt}]$ [1], with $\alpha = g, e$, the projection operators $P_j^0 = 1 - \hat{\sigma}_j^{rr}$ and $P_j^1 = \hat{\sigma}_j^{rr}$. After transformation the Hamiltonian reads

$$\hat{H}(t) = \sum_j \sum_k \left( \frac{\theta_j^k}{2} P_{j-1}^0 \hat{\sigma}_j^{gr} P_{j+1}^0 + \frac{\phi_j^k}{2} P_{j-1}^0 \hat{\sigma}_j^{er} P_{j+1}^0 \right) e^{ikVt}$$
$$+ \left( \frac{\theta_j^k}{2} P_{j-1}^0 \hat{\sigma}_j^{gr} P_{j+1}^1 + \frac{\phi_j^k}{2} P_{j-1}^0 \hat{\sigma}_j^{er} P_{j+1}^1 \right) e^{i(k-1)Vt}$$
$$+ \left( \frac{\theta_j^k}{2} P_{j-1}^1 \hat{\sigma}_j^{gr} P_{j+1}^0 + \frac{\phi_j^k}{2} P_{j-1}^1 \hat{\sigma}_j^{er} P_{j+1}^0 \right) e^{i(k-1)Vt}$$
$$+ \left( \frac{\theta_j^k}{2} P_{j-1}^1 \hat{\sigma}_j^{gr} P_{j+1}^1 + \frac{\phi_j^k}{2} P_{j-1}^1 \hat{\sigma}_j^{er} P_{j+1}^1 \right) e^{i(k-2)Vt}$$
$$+ h.c. \qquad (5)$$

This can be written as

$$\hat{H}(t) = \sum_j \sum_{k,k'} \hat{H}_{k,k'}(j) e^{i(k-k')Vt} + h.c. \qquad (6)$$

where $k, k' = \{0 \ldots 2\}$ and the non-hermitian time-independent operators $\hat{H}_{k,k'}$ are given by

$$\hat{H}_{k,0}(j) = \frac{\theta^k}{2} P_{j-1}^0 \hat{\sigma}_j^{gr} P_{j+1}^0 + \frac{\phi^k}{2} P_{j-1}^0 \hat{\sigma}_j^{er} P_{j+1}^0 \qquad (7)$$

$$\hat{H}_{k,1}(j) = \frac{\theta^k}{2} P_{j-1}^0 \hat{\sigma}_j^{gr} P_{j+1}^1 + \frac{\phi^k}{2} P_{j-1}^0 \hat{\sigma}_j^{er} P_{j+1}^1$$
$$\qquad + \frac{\theta^k}{2} P_{j-1}^1 \hat{\sigma}_j^{gr} P_{j+1}^0 + \frac{\phi^k}{2} P_{j-1}^1 \hat{\sigma}_j^{er} P_{j+1}^0$$

$$\hat{H}_{k,2}(j) = \frac{\theta^k}{2} P_{j-1}^1 \hat{\sigma}_j^{gr} P_{j+1}^1 + \frac{\phi^k}{2} P_{j-1}^1 \hat{\sigma}_j^{er} P_{j+1}^1$$

Next we approximate Eq. (6) using the formalism described in [Ref. [2] Eq.(35)]. This approach is equivalent to second order adiabatic elimination in the Floquet picture. The effective Hamiltonian [3] reads

$$\hat{H}' = \sum_k (\hat{H}_{k,k} + \hat{H}_{k,k}^\dagger) + \frac{1}{V} \sum_{k' \neq k} \frac{1}{k' - k} [\hat{H}_{k,k'}^\dagger, \hat{H}_{k,k'}] \qquad (8)$$

where the first term describes the resonant couplings and the last term includes leading order corrections due to off-resonant cross-talk couplings. These terms enter as

effective light shifts and couplings between states within constant $k$ manifolds (scaling with $(\theta^k)^2/V$ and $(\phi^k)^2/V$). For the manuscript we restrict to the resonant terms only. Thus one can write the time-independent Hamiltonian as

$$\hat{H}' = \frac{1}{2} \sum_j \sum_{\alpha,\beta} \mathrm{P}_{j-1}^\alpha \left[\theta_j^k \hat{\sigma}_j^{gr} + \phi_j^k \hat{\sigma}_j^{er} + h.c.\right] \mathrm{P}_{j+1}^\beta \quad (9)$$

where we introduce the projection indices $\alpha, \beta = \{0, 1\}$, with $k = \alpha + \beta$.

Finally we include the spontaneous decay of the $|e\rangle$ state and derive an effective master equation for the dynamics of the form $\partial_t \rho = -i[\hat{H}_{\rm eff}, \rho] + \mathcal{D}_{\hat{L}^{\rm eff}}[\rho]$. Spontaneous emission from the $|e\rangle$ state to the $|g\rangle$ state with rate $\Gamma$ is described by the jump operators $\hat{L}_j = \sqrt{\Gamma}\hat{\sigma}_j^{ge}$. To adiabatically eliminate the manifold containing short lived $|e\rangle$ states and to derive effective jump operators $L_j^{\rm eff}$ acting on the $g, r$ subspace we use the effective operator formalism of [4].

We start by defining operators for the slow and fast evolving subspaces according to $\mathbb{P}|e_j\rangle = 0$ and $\mathbb{Q} = 1 - \mathbb{P}$, such that $\mathbb{Q}\sigma_j^{e\alpha} = \sigma_j^{e\alpha}$ and $\mathbb{P}\sigma_j^{e\alpha} = 0$ ($\alpha = g, r$). Following Ref. [4] we also define the non-hermitian Hamiltonian

$$\hat{H}_{nh} = \mathbb{Q}\hat{H}'\mathbb{Q} - \frac{i}{2}\sum_j \hat{L}_j^\dagger \hat{L}_j. \quad (10)$$

As a first approximation we neglect states with more than one $|e\rangle$ excitation which is justified for $\Gamma \gg |\phi_k|$. In this case $\mathbb{Q}\hat{H}'\mathbb{Q} = 0$, which leaves

$$\hat{H}_{nh} = -\frac{i\Gamma}{2}\sum_j \hat{\sigma}_j^{ee}. \quad (11)$$

The effective Hamiltonian can be written as

$$\hat{H}_{\rm eff} = \mathbb{P}\hat{H}'\mathbb{P} - \frac{1}{2}\mathbb{V}^-[\hat{H}_{nh}^{-1} + (\hat{H}_{nh}^{-1})^\dagger]\mathbb{V}^+ \quad (12)$$

where $\mathbb{V}^- = \mathbb{P}\hat{H}'\mathbb{Q}$ and $\mathbb{V}^+ = \mathbb{Q}\hat{H}'\mathbb{P}$. Since the inverse of $\hat{H}_{nh}$ is a purely imaginary diagonal matrix the term in the brackets cancels, leaving

$$\hat{H}_{\rm eff} = \mathbb{P}\hat{H}'\mathbb{P}$$
$$= \frac{1}{2}\sum_j \sum_{\alpha,\beta} \mathrm{P}_{j-1}^\alpha \left[\theta_j^k \hat{\sigma}_j^{gr} + (\theta_j^k)^* \hat{\sigma}_j^{rg}\right] \mathrm{P}_{j+1}^\beta, \quad (13)$$

which coincides with Eq. (2) in the manuscript.

The effective jump operators can be written

$$\hat{L}_j^{\rm eff} = \hat{L}_j \hat{H}_{nh}^{-1} \mathbb{V}^+$$
$$= \frac{i}{\sqrt{\Gamma}} \hat{\sigma}_j^{ge}(\sum_i \hat{\sigma}_i^{ee})^{-1}$$
$$\times \mathbb{Q}\sum_{i'}\left(\sum_{\alpha,\beta}\mathrm{P}_{i'-1}^\alpha[\theta_j^k\hat{\sigma}_{i'}^{gr}+\phi_j^k\hat{\sigma}_{i'}^{er}+h.c]\mathrm{P}_{i'+1}^\beta\right)\mathbb{P}.$$
$$\quad (14)$$

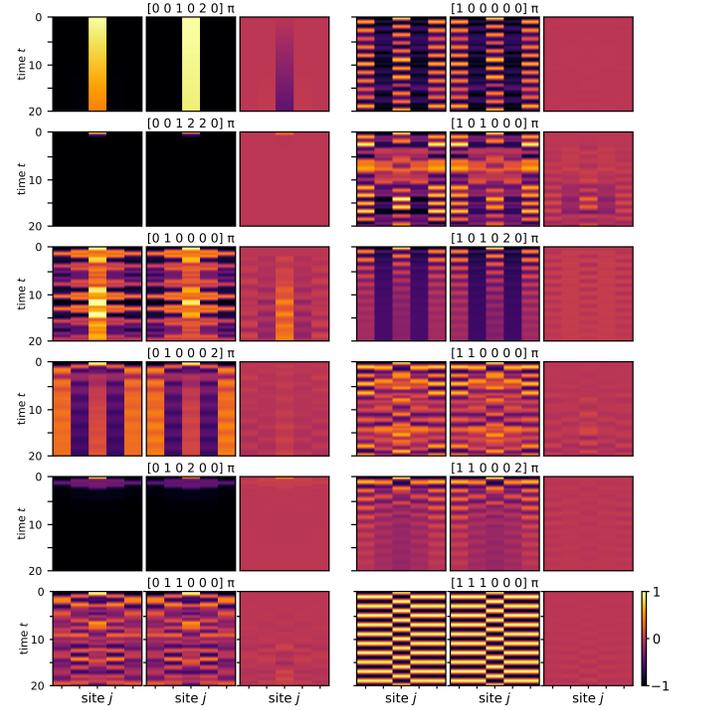

Figure S1. Numerical simulations of the evolution of the magnetization $\langle \hat{Z}_j \rangle$ for the full three-level model as well as its effective two-level description for 12 representative QCA rules indicated in the label of each subfigure with the notation $[\theta^0, \theta^1, \theta^2, \tilde{\phi}^0, \tilde{\phi}^1, \tilde{\phi}^2]$. For each QCA rule there are three panels: *left:* three-level master equation simulation including time-dependent couplings, *center:* two-level effective master equation simulation and *right:* residual between three-level and two-level result.

Now we use $(\sum \hat{\sigma}^{ee})^{-1}\mathbb{Q} = (\sum \hat{\sigma}^{ee})^{-1}$ and the action of $\mathbb{P}$ from the right hand side changes $\mathrm{P}_{j\pm 1}^0 \to |g\rangle\langle g|_{j\pm 1}$ and restricts $\hat{\sigma}_{i'}^{er}$ to states that initially have no $|e\rangle$ excitations (i.e. exactly one $|e\rangle$ at site $i'$ after application of the operator). This allows us to remove the sums over $i, i'$

$$\hat{L}_j^{\rm eff} = \frac{i}{\sqrt{\Gamma}}\sum_{\alpha,\beta}\mathrm{P}_{j-1}^\alpha \phi_j^k \hat{\sigma}_j^{ge}(\hat{\sigma}_j^{ee})^{-1}\hat{\sigma}_j^{er}\mathrm{P}_{j+1}^\beta \quad (15)$$
$$= \frac{i}{\sqrt{\Gamma}}\sum_{\alpha,\beta}\phi_j^k \mathrm{P}_{j-1}^\alpha \hat{\sigma}_j^{gr}\mathrm{P}_{j+1}^\beta.$$

To arrive at Eq. (3) in the manuscript we drop the $i$ prefactor, since it is of no physical consequence [canceling out in the Lindblad term $\mathcal{D}[\rho] = \sum_j \hat{L}_j \rho \hat{L}_j^\dagger - (\hat{L}_j^\dagger \hat{L}_j \rho + \rho \hat{L}_j^\dagger \hat{L}_j)/2]$.

# NUMERICAL COMPARISON BETWEEN THE THREE-LEVEL MULTI-FREQUENCY HAMILTONIAN AND THE EFFECTIVE TWO-LEVEL MODEL

To verify the validity of the effective two-level model we compare numerical simulations of the effective model and the full three-level Rydberg excitation Hamiltonian (including time-dependent couplings). Fig. S1 shows the time evolution of the magnetization $\langle \hat{Z}_j \rangle$ for the same 12 representative rules that were chosen in Fig. 2 in the manuscript for 5 sites with open boundary conditions. For each rule we display three panels: on the left is the result of the three-level master equation [Eq.(1) in the manuscript]; in the middle is the effective two-level master equation results [Eqs. (2) and (3) in the manuscript]; and the right panels show the difference between the three-level and two-level results, where more homogeneous colors indicate better agreement. The simulations are performed with with a nearest neighbor Rydberg-Rydberg interaction energy $V = 50\pi$ (in units where $t = 1$), an intermediate state decay rate of $\Gamma = 6\pi$ and coupling parameters $\theta^k, \tilde{\phi}^k$ indicated in the figure labels, with $\phi^k = \sqrt{\Gamma \tilde{\phi}^k}$. We additionally include jump operators for Rydberg state decay with the dimensionless rate $\gamma = 8\pi \cdot 10^{-4}$. Each rule is evolved starting in the initial state $|00100\rangle$ for a duration of 20 time units.

Inspecting the results in Fig. S1 we see good agreement between the effective model and the full three-level time-dependent master equation for the majority of rules. This agreement is especially good considering the the chosen parameters are close to the limit of validity for the approximations used in deriving the effective model (i.e. $V \gg \Gamma \gg \theta^k, \phi^k \gg \gamma$). Certain rules do show some deviations though, e.g. rule $[0, 0, 1, 0, 2, 0] \times \pi$, where the excited state population of the central site in the three-level case is seen to decay with a time constant of around 50 time units, which is not captured by the effective two-level model. This can be attributed to slow off-resonant depumping caused by the $\phi^{k=1}$ coupling acting on the $k = 0$ subspace which is neglected in the effective model. Increasing $V/\Gamma$ further improves the agreement between the two models.

# VARIATIONAL STATE PREPARATION OF THE ANTIFERROMAGNETIC GHZ STATE

In this section we demonstrate the versatility of the QCA approach to engineer different quantum states. In particular we show how *minimizing* $\langle C \rangle = 1/N \sum_j C_{j,j+1}$ using variational optimization realizes the *antiferromagnetic* GHZ state. We otherwise use the same parameters as in the main manuscript: we start from initial state $|0\rangle^{\otimes N} = |000000\rangle$ and use particle swarm optimization (PSO) to find the set of parameters $[\theta^0, \theta^1, \theta^2, \tilde{\phi}^0, \tilde{\phi}^1, \tilde{\phi}^2]$,

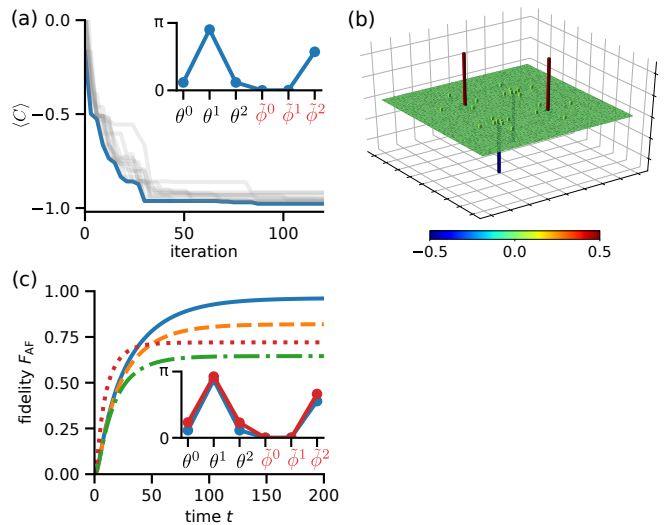

Figure S2. Variational quantum optimization of the antiferromagnetic GHZ state for a chain of $N = 6$ sites. (a) Convergence of particle swarm optimization with a population size of 10 (gray lines) towards $\langle C \rangle \approx -1$. The solid blue line highlights the best individual. The optimal parameters are shown in the inset (blue points). (b) Graphical representation of the density matrix of the optimized state (imaginary parts are $< 6 \cdot 10^{-4}$). (c) Time evolution of the fidelity between the resulting QCA state using the optimal variational parameters and the antiferromagnetic $\mathrm{GHZ}^N$ state. The fidelity exceeds the classical threshold of 0.5 within approximately 20 time units, and saturates to a value close to 1 after 100-200 time units. The dashed orange (dashed-dotted green) line shows the same evolution including a Rydberg state decay rate of $\gamma/2\pi = 0.8\,\mathrm{kHz}$ ($\gamma/2\pi = 2.4\,\mathrm{kHz}$). Reoptimization of the QCA parameters including Rydberg state decay with $\gamma/2\pi = 2.4\,\mathrm{kHz}$ gives slightly different parameters (red points in the inset) and improves the attainable fidelity by approximately 10% (see the text for other parameters).

that minimizes $\langle C \rangle$. Fig. S2(a) displays the convergence during the optimization loop for a population of size 10 (gray lines) towards small covariance coefficients $\langle C \rangle$. The solid blue line depicts the best individual of all populations. The optimal parameters are $\pi \times [0.122, 0.941, 0.122, 0., 0.003, 0.597]$, shown as inset in Fig. S2(a). This can be compared to the results for the *ferromagnetic* GHZ state presented in the main paper which had a comparable convergence time, but different set of optimal parameters: $\pi \times [0.058, 0.938, 0.058, 0.626, 0., 0.]$ (displayed also in Fig. 3(b) inset).

The density matrix of the resulting state resembles closely the density matrix of the 6 atom antiferromagnetic GHZ state $|\mathrm{AF}\rangle = 1/\sqrt{2}(|010101\rangle - |101010\rangle)$, Fig. S2(b). This can be understood, as the two states $|010101\rangle$ and $|101010\rangle$ are dark with respect to the projectors associated with two main components $\theta^1$ and $\tilde{\phi}^2$. Fig. S2(c) displays the fidelity with the antiferromagetic GHZ state during time evolution using the found optimal parameters. Calculations that neglecting uncontrolled dissipation (decay

of the Rydberg state) show that the classical threshold corresponding to a fidelity of 0.5 is surpassed after 20 time steps (blue curve), reaching a final value close to 1 after $t \sim 150$ time units. The inclusion of additional jump operators describing Rydberg state decay with a rate of $\gamma/2\pi = 0.8\,\text{kHz}$ (2.4 kHz) reduces the fidelity $F \gtrsim 0.82$ (0.65), shown as orange dashed (dotted-dashed green) line.

Finally, we show that the variational optimization procedure can improve the generation of the antiferromagnetic GHZ state in the presence of uncontrolled dissipation. Optimizing the QCA parameters while including additional Rydberg state decay with $\gamma/2\pi = 2.4\,\text{kHz}$ yields slightly different optimum parameter values of $\pi \times [0.251, 0.998, 0.248, 0., 0., 0.715]$, shown as red points in Fig. S2(c-inset), as compared to the case with $\gamma = 0$ (blue points). The approach to the stationary state is shown as a red dotted red curve in Fig. S2(c). Both the rate at which the system reaches its stationary value and the maximum fidelity $F \gtrsim 0.72$ is increased as compared to the case using the $\gamma = 0$ parameters shown in green.

---

[1] I. Lesanovsky, "Many-body spin interactions and the ground state of a dense Rydberg lattice gas," Phys. Rev. Lett **106**, 025301 (2011).

[2] M. Sanz, E. Solano, and Í. L. Egusquiza, "Beyond adiabatic elimination: Effective hamiltonians and singular perturbation," in *Applications + Practical Conceptualization + Mathematics = fruitful Innovation*, edited by Robert S. Anderssen, Philip Broadbridge, Yasuhide Fukumoto, Kenji Kajiwara, Tsuyoshi Takagi, Evgeny Verbitskiy, and Masato Wakayama (Springer Japan, Tokyo, 2016) pp. 127–142.

[3] D. F. James and J. Jerke, "Effective Hamiltonian theory and its applications in quantum information," Can. J. Phys **85**, 625–632 (2007).

[4] F. Reiter and A. S. Sørensen, "Effective operator formalism for open quantum systems," Phys. Rev. A **85**, 032111 (2012).

 

\end{document}